%% file: main.tex
\documentclass{article}
\usepackage[main, final]{neurips_2025}
\usepackage[utf8]{inputenc} % allow utf-8 input
\usepackage[T1]{fontenc}    % use 8-bit T1 fonts
\usepackage{hyperref}       % hyperlinks
\usepackage{url}            % simple URL typesetting
\usepackage{booktabs}       % professional-quality tables
\usepackage{amsfonts}       % blackboard math symbols
\usepackage{nicefrac}       % compact symbols for 1/2, etc.
\usepackage{microtype}      % microtypography
\usepackage{xcolor}         % colors
\usepackage{graphicx}
\usepackage{adjustbox}
\usepackage{enumitem}
\usepackage{amsmath,amsfonts}
\usepackage{algorithmic}
\usepackage{array}
\usepackage[caption=false,font=normalsize,labelfont=sf,textfont=sf]{subfig}
\usepackage{textcomp}
\usepackage{stfloats}
\usepackage{verbatim}
\usepackage{balance}
\usepackage{multirow}
\usepackage{threeparttable}
\usepackage{arydshln}
\usepackage{soul}
\usepackage{fontawesome}
\usepackage{makecell}
\usepackage{natbib}
\usepackage{siunitx}
\usepackage{colortbl}
\usepackage{tikz}
\usepackage{aecompl}
\usepackage{xparse}

\usepackage{xspace}
\usepackage{tabularx}
\usepackage{wrapfig}
\usepackage{subcaption}
\usepackage{rotating}
\usepackage{multicol}
\usepackage{setspace}
\usepackage{adjustbox} 
\usepackage{tipa}
\usepackage[most]{tcolorbox}
\definecolor{mygray}{gray}{0.9}
\definecolor{mydarkblue}{rgb}{0,0.08,0.45}
\definecolor{mylightblue}{rgb}{0,1,1}
\hypersetup{
 colorlinks=true,
 citecolor=mydarkblue,
 urlcolor=mydarkblue,
 linkcolor=black,
}
\usepackage{footmisc}

\usepackage{eso-pic} % provides AddToShipoutPicture
\usetikzlibrary{calc} % <-- add this
\AddToShipoutPictureBG*{%
  \begin{tikzpicture}[remember picture,overlay]
        \node[anchor=north east] at ($(current page.north east)+(-2.95cm,-0.3cm)$)
      {\includegraphics[width=5cm]{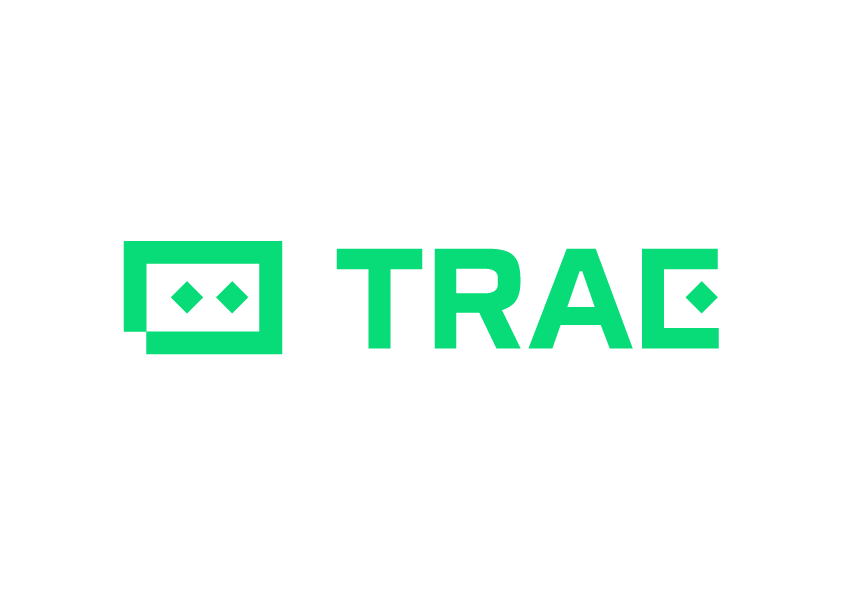}};
  \end{tikzpicture}%
}

\newcommand{\tech}{\texttt{Trae Agent}}

\newcommand{\Comment}[1]{}

\title{Trae Agent: An LLM-based Agent for Software Engineering with Test-time Scaling}
\author{%
Trae Research\\
Beijing, China\\
\href{mailto:opensource@mail.trae.ai}{\texttt{opensource@mail.trae.ai}}
}

\begin{document}

\maketitle

\renewcommand{\thefootnote}{}
\footnotetext{
\hyperref[sec:contributors]{The list of authors is provided at the end of the technical report.}
}

\begin{abstract}
Software issue resolution is a critical challenge in software engineering and has garnered increasing attention in recent years. 
With the rapid advancement of large language models (LLMs), substantial progress has been made in addressing real-world software engineering tasks.
Recent studies have introduced ensemble reasoning techniques to enhance the performance of LLM-based issue resolution.
However, existing prompting-based methods still face limitations in effectively exploring large ensemble spaces and lack the capacity for repository-level understanding, both of which constrain their overall effectiveness.
In this paper, we propose \tech{}, the first agent-based ensemble reasoning approach for repository-level issue resolution.
\tech{} formulates our goal as an optimal solution search problem and addresses two key challenges, i.e., large ensemble spaces and repository-level understanding, through modular agents for generation, pruning, and selection.
We conduct extensive experiments using three leading LLMs on the widely-adopted SWE-bench benchmark, comparing \tech{} against four state-of-the-art ensemble reasoning techniques. 
Experimental results demonstrate that \tech{} consistently achieves superior performance, with an average improvement of 10.22\% over all baselines in terms of Pass@1.
\tech{} \textit{has achieved first place on the SWE-bench Verified leaderboard, with a notable Pass@1 score of 75.20\%.}
We are pleased to release \tech{} as an open-source project to support the research community, with all resources available at \url{https://github.com/bytedance/trae-agent}.
\end{abstract}

\input{introduction}

\input{motivation}
\input{approach}
\input{evaluation}
\input{results}

\input{discussion}
\input{threats}
\input{related}

\section{Conclusion}
\label{sec:conclusion}
In this work, we present \tech{}, the first agent-based ensemble reasoning approach for repository-level issue resolution, which substantially improves the effectiveness of LLMs.
\tech{} formulates ensemble reasoning as an optimal solution search problem and tackles two fundamental challenges, i.e., large ensemble spaces and repository-level understanding, through modular agents for generation, pruning, and selection.
To comprehensively evaluate its performance, we conduct extensive experiments using three leading LLMs on the widely-used SWE-bench benchmark. 
The experimental results show that \tech{} consistently outperforms four state-of-the-art ensemble reasoning baselines across all evaluation settings, demonstrating its effectiveness in enhancing LLM-based software issue resolution.

\input{appendix}

\bibliographystyle{abbrvnat}
\bibliography{reference}

\end{document}

%% file: introduction.tex
\section{Introduction}
\label{sec:introduction}
Software issue resolution refers to the automated handling of newly reported bugs or feature requests during software development and maintenance, aiming to ensure correct and reliable system behavior~\citep{fakhoury2023towards,jimenezswe,guo2025omnigirl}.
In recent years, this task has garnered growing interest from both academia and industry, driven by its potential to reduce developer burden and significantly enhance productivity~\citep{yang2024swe,zhang2024autocoderover}.
Large language models (LLMs) have shown remarkable capabilities in function-level code tasks such as code generation~\citep{lin2025flowgen,tian2025fixing} and automated program repair~\citep{fan2023automated,bouzenia2025repairagent}.
Despite this progress, LLMs continue to face major challenges in resolving complex repository-level software issues.
For example, GPT-4o achieves a resolution rate of 92.70\% on the function-level HumanEval benchmark~\citep{liu2023your}, but only 11.99\% on the repository-level SWE-bench benchmark~\citep{yang2024swe}.
This performance gap underscores the difficulty of real-world software issue resolution, which often requires global understanding of large codebases, cross-file reasoning, and the detection of subtle, multi-component bugs~\citep{ruan2025specrover,xia2024agentless}.
These challenges limit the practical adoption of LLM-based techniques in real-world software engineering and pose a risk to software quality if unresolved.
Bridging this gap remains a critical research goal toward realizing robust, scalable, and reliable automated software issue resolution.

Recently, many techniques have been proposed to enhance the performance of LLMs in software issue resolution~\citep{yang2024swe,wang2025openhands,zhang2024autocoderover}.
These methods primarily focus on improving patch generation, by carefully designing agent architectures and integrating external tools to assist LLMs in reasoning and producing correct fixes.
For example, OpenDevin\citep{wang2025openhands} incorporates a planning module based on issue descriptions and coordinates multiple tools within an agent system to guide patch generation.
While these approaches improve individual patch generation, recent studies~\citep{zhang2024diversity,li2025s,augmenttop12025} have revealed a key property of LLMs: although the overall resolution rate remains stable across repeated runs, the set of issues successfully resolved varies across executions.
This variability arises from the large action space and complex reasoning trajectories required for repository-level issue resolution, which often lead LLMs to pursue divergent solution paths.
This observation has led to the emergence of a new direction, i.e., \textbf{ensemble reasoning}, which builds on patch generation approaches by producing multiple independent candidate patches and selecting a consensus solution that is more robust than any individual output.

Existing ensemble reasoning methods are prompting-based, relying on LLMs not only to generate candidate patches, but also to perform selection among them.
For example, Augment\citep{augmenttop12025} uses the LLM-as-a-judge~\citep{wang2025can} paradigm to prompt the model to compare each candidate against the issue description and choose the best match.
DeiBase\citep{zhang2024diversity} prompts the LLM to provide justification and confidence scores for each candidate patch, and selects the highest-scoring one.
While prompting-based ensemble reasoning techniques have shown promise, they face fundamental limitations when applied to complex repository-level issue resolution.
These limitations become especially evident in the face of two key challenges:
First, prompting-based methods \textit{struggle with the optimal solution search in large ensemble spaces}.
As the number of candidate patches grows, subtle behavioral differences between them become increasingly difficult to distinguish using LLMs alone.
Since patch selection is typically performed through a single prompt, surface-level syntactic comparisons often fail to capture deeper semantic nuances.
Second, these methods \textit{lack the capacity for repository-level understanding}, which is essential for accurately selecting patches in real-world software systems.
Issues and fixes frequently span multiple files and modules, requiring cross-file reasoning, contextual awareness, and verification of patch correctness within the broader codebase.
Prompting-based approaches operate in a stateless, single-turn manner and lack persistent memory or tool integration, making it difficult to track dependencies, execute validation steps, or build a coherent global view of the repository.

To overcome the limitations of prompting-based ensemble reasoning approaches, we propose the first agent-based ensemble reasoning approach for repository-level issue resolution, called \textbf{\tech{}}. 
This approach is designed to enhance LLM-based issue resolution by formulating it as an optimal solution search problem~\citep{harman2001search,clarke2003reformulating} and addressing the two core challenges through a modular agent-based architecture.
It consists of three key components: \textit{patch generation}, \textit{patch pruning}, and \textit{patch selection}. 
To improve the ensemble diversity, \tech{} leverages a novel coder agent that first generates diverse candidate patches in parallel.
To mitigate the difficulty of optimal solution search, \tech{} applies a hierarchical pruning strategy that combines patch deduplication and regression testing strategies to eliminate redundant or faulty patches, thereby reducing the ensemble space while preserving promising candidates. 
To address the need for repository-level understanding, \tech{} employs a selector agent that simulates a real-world program comprehension process~\citep{eisenbarth2001aiding,di2005integrating}, encompassing both static review and dynamic verification. 
Specifically, this agent iteratively gathers and analyzes relevant code snippets (such as those referenced in the issue description, modified by patches, or related through dependencies) to build static understanding, and collects execution traces from automatically generated tests to build dynamic understanding.
Based on these insights, the agent applies a majority voting strategy to select the most plausible patch. 
Due to its generalizable and modular design, \tech{} potentially provides a solid foundation for advancing ensemble reasoning in other complex software engineering tasks.

We conduct extensive experiments to evaluate the performance of \tech{} using three state-of-the-art LLMs (i.e., Gemini 2.5 Pro~\citep{gemini25pro2025}, Claude 3.7 Sonnet~\citep{claude37sonnet2025}, and GPT-4.1~\citep{gpt412025}) on the widely-used software issue resolution benchmark (i.e., SWE-bench~\citep{jimenezswe}). 
Our results show that \tech{} consistently and significantly outperforms four state-of-the-art ensemble reasoning baselines (i.e., Augment, Augment w/ Pruning, DeiBase, and DeiBase w/ Pruning) across all evaluation settings, demonstrating its capability to enhance LLM-based software issue resolution.
In particular, \tech{} achieves a Pass@1 improvement ranging from 5.83\% to 14.60\%, where Pass@1 denotes the proportion of generated patches that pass all golden tests.
We further investigate the impact of ensemble size, a critical hyper-parameter in ensemble reasoning.
The results reveal that \tech{} not only outperforms all baselines at each ensemble size but also continues to improve with larger ensemble sizes, while baseline methods exhibit varying degrees of performance degradation.
Additionally, we conduct comprehensive ablation studies by constructing five variants of \tech{}, each omitting a major component (i.e., patch pruning, patch deduplication, regression testing, selector agent, and majority voting). 
The results confirm the effectiveness and necessity of each part of our framework.

The main contributions of this paper are as follows:
\begin{itemize}[leftmargin=10pt]
    \item \textbf{Novel Approach}: 
    We propose \tech{}, the first agent-based ensemble reasoning framework for repository-level issue resolution. 
    \tech{} formulates our goal as an optimal solution search problem and addresses two key challenges, i.e., large ensemble spaces and repository-level understanding, through modular agents for generation, pruning, and selection.
    
    \item \textbf{Extensive Evaluation}: 
    We conduct extensive experiments on the SWE-bench benchmark using three state-of-the-art LLMs, comparing \tech{} with four leading baselines. 
    Results show that \tech{} consistently improves Pass@1 by 5.83\%$\sim$14.60\%, demonstrating its robust effectiveness across settings.
    Particularly, \tech{} has achieved first place on the SWE-bench Verified leaderboard, with a notable Pass@1 score of 75.20\%.
    
    \item \textbf{Public Artifact}: 
    To support the research community and facilitate future research, we release all resources available at \url{https://github.com/bytedance/trae-agent}.
    Our open-source GitHub repository has attracted over 8,000 stars (as of July 2025), reflecting strong community interest and adoption.
\end{itemize}

%% file: motivation.tex
\section{Motivation}
\label{sec:motivation}

\begin{figure*}[t!]
    \centering
    \includegraphics[width=0.9\linewidth]{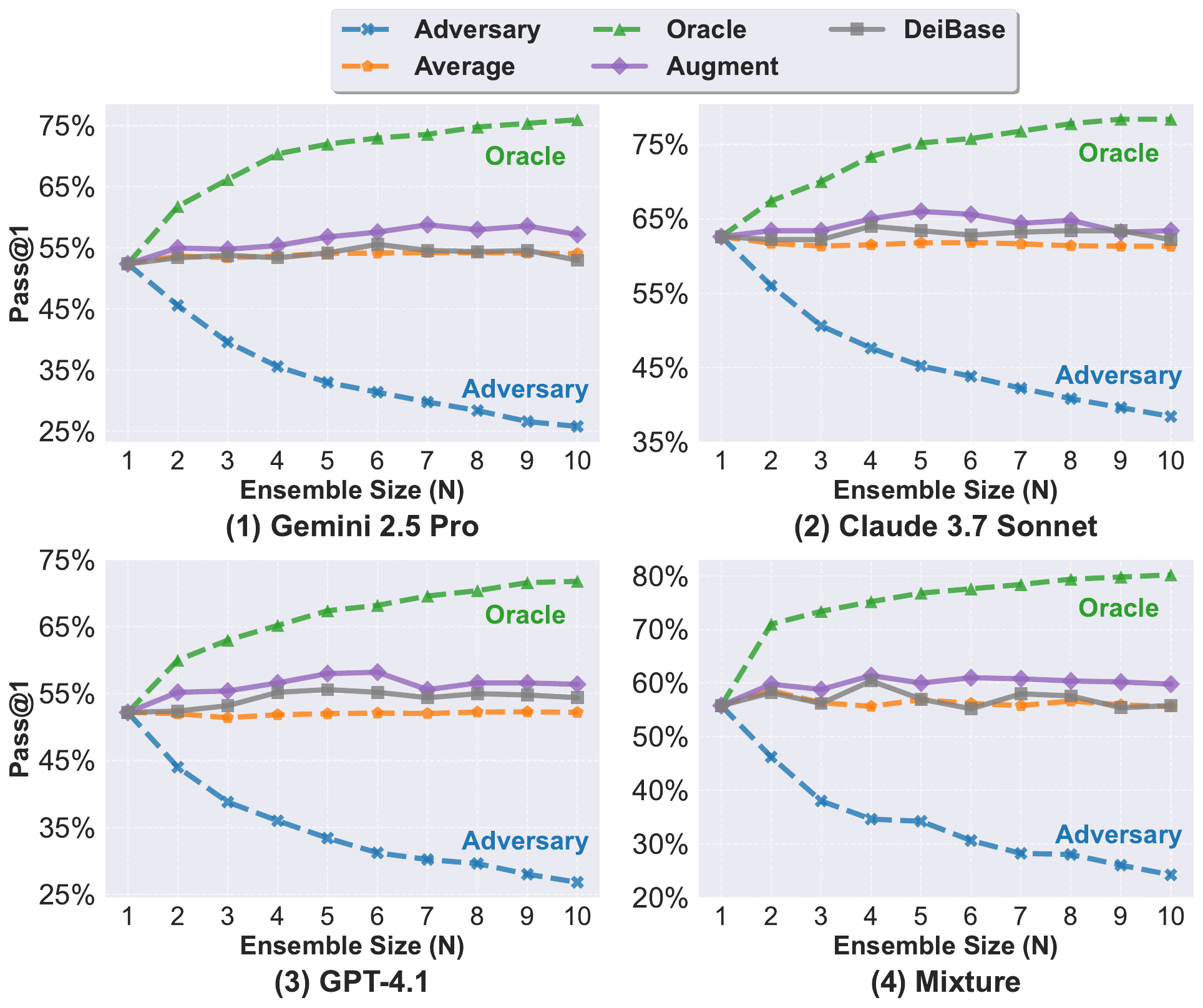}
    \caption{Influence of ensemble size on the effectiveness of existing ensemble reasoning techniques in terms of Pass@1 ($\uparrow$)}
    \label{fig:motivation}
\end{figure*}

To motivate the potential of ensemble reasoning approaches and the necessity of the two key components in \tech{}, i.e., patch pruning (Section~\ref{subsec:pruning}) and patch selection (Section~\ref{subsec:selection}), we present a comprehensive empirical analysis.

Figure~\ref{fig:motivation} shows the performance of two state-of-the-art ensemble reasoning techniques (Augment~\citep{augmentagent2025} and DeiBase~\citep{zhang2024diversity}) on the widely-used SWE-bench Verified~\citep{jimenezswe} benchmark using three leading LLMs (Gemini 2.5 Pro~\citep{gemini25pro2025}, Claude 3.7 Sonnet~\citep{claude37sonnet2025}, and GPT-4.1~\citep{gpt412025}).
In particular, the Mixture setting refers to a round-robin combination of these three LLMs for generating candidate patches, aiming to enhance the diversity of the ensemble space.
Taking Figure~\ref{fig:motivation}(1) as an example, the x-axis denotes the ensemble size $N$ (i.e., the number of candidate patches generated by Gemini 2.5 Pro), while the y-axis shows the Pass@1 (i.e., the proportion of patches that pass all golden tests) metric.
Following prior work~\citep{zhang2024diversity}, we report three additional reference metrics: 
(1) \textit{Oracle}, representing the best-case performance where the correct patch is always selected if present among the $N$ candidates; 
(2) \textit{Adversary}, representing the worst-case performance where an issue is only considered solved if all $N$ candidates are correct; 
and (3) \textit{Average}, representing the expected performance of randomly selecting one patch from the $N$ candidates.

First, we observe that as the ensemble size increases, the performance of \textit{Average} remains relatively stable, while the \textit{Oracle} value consistently improves. 
Notably, when the ensemble size reaches 10, \textit{Oracle} achieves an average improvement of 20.80\% over \textit{Average} in terms of Pass@1. 
This observation strongly supports previous findings that although the overall resolution rate remains stable across repeated runs, the set of issues successfully resolved varies across executions~\citep{zhang2024diversity,li2025s,augmenttop12025}.
This finding reflects the high diversity in the candidate patches generated by LLMs. 
\textit{It motivates the significant potential of ensemble reasoning techniques as a promising direction for enhancing LLM-based software issue resolution.}

Moreover, we observe that as the ensemble size increases, the \textit{Adversary} value consistently declines. 
Specifically, when the ensemble size reaches 10, \textit{Adversary} reduces by 27.01\% on average compared to the \textit{Average} in terms of Pass@1. 
This indicates that although the theoretical upper bound of ensemble performance (i.e., \textit{Oracle} = 20.80\%) improves as the number of candidate patches increases, the theoretical lower bound (i.e., \textit{Adversary} = 27.01\%) deteriorates more significantly.
This widening gap underscores the increasing difficulty of ensemble reasoning techniques as ensemble size increases.
Indeed, as shown in Figure~\ref{fig:motivation}, the performance of Augment and DeiBase initially improves with larger ensemble sizes but subsequently degrades. 
\textit{It motivates the necessity of our patch pruning component, which eliminates redundant and faulty patches, thereby reducing the ensemble space while preserving promising candidates.}

Additionally, on average, both Augment and DeiBase outperform the random selection strategy (i.e., \textit{Average}) by 3.17\% and 1.06\% in terms of Pass@1, respectively.
These results also confirm the potential of ensemble reasoning techniques to enhance the LLM performance in software issue resolution. 
However, in comparison, \textit{Oracle} significantly achieves the average improvement of 29.49\% and 36.29\% over Augment and DeiBase, respectively. 
This substantial performance gap underscores the limitations of existing ensemble reasoning methods in accurately selecting the correct patch and highlights significant potential for further improvement.
\textit{It motivates the necessity of designing a more effective patch selection component to further improve the LLM performance in software issue resolution.}

%% file: approach.tex
\section{Approach}
\label{sec:approach}
\subsection{Overview}
\label{subsec:overview}

\begin{figure*}[t!]
    \centering
    \includegraphics[width=1.0\linewidth]{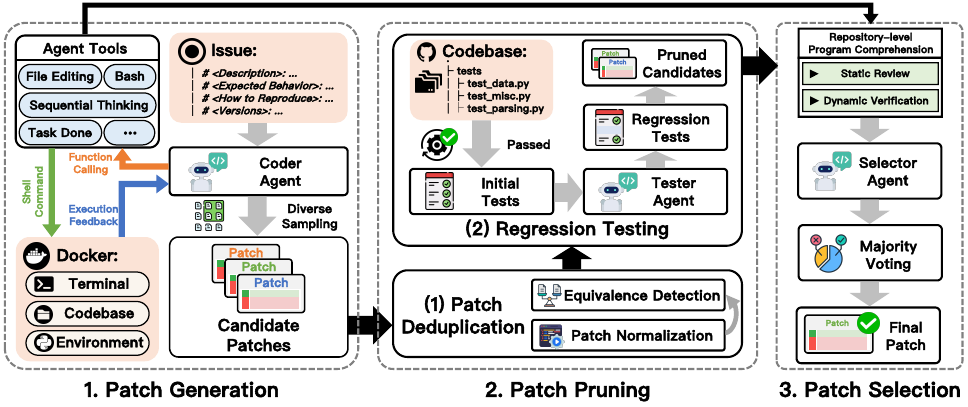}
    \caption{Overview of Trae Agent}
    \label{fig:overview}
\end{figure*}

\textit{Problem setup}:
Given a software codebase $C$ and a natural-language-described software issue $I$ (which may involve a bug fix or feature implementation), the goal of an ensemble reasoning approach is to automatically generate a diverse set of candidate patches $P = \{p_1, p_2, ..., p_N\}$ and then select a consensus patch $p_i$ ($1 \leq i \leq N$) to resolve $I$. 
Applying $p_i$ to the original codebase $C$ yields a modified codebase $C'$. 
The selected patch $p_i$ is considered correct if $C'$ passes all golden tests $T$ and satisfies all requirements specified in $I$.

In this paper, we propose a novel ensemble reasoning approach, \tech{}, designed to enhance the performance of LLMs in software issue resolution. 
To the best of our knowledge, \tech{} is the first agent-based ensemble reasoning approach for repository-level issue resolution. 
Due to its generalizable and modular design, \tech{} potentially provides a solid foundation for advancing ensemble reasoning in other complex software engineering tasks.
Figure~\ref{fig:overview} illustrates the overall architecture of \tech{}, which comprises three main components:
\begin{itemize}[leftmargin=10pt]
    \item \textbf{Patch generation component}(Section~\ref{subsec:generation}) employs a novel coder agent to generate diverse candidate patches in parallel to improve the ensemble diversity;
    \item \textbf{Patch pruning component} (Section~\ref{subsec:pruning}) performs a hierarchical patch pruning that combines patch deduplication and regression testing strategies to eliminate redundant or faulty patches, thereby reducing the ensemble space while preserving promising candidates;
    \item \textbf{Patch selection component} (Section~\ref{subsec:selection}) employs a selector agent that simulates a real-world program comprehension process~\citep{eisenbarth2001aiding,di2005integrating}, constructing repository-level understanding by combining static review with dynamic verification, and ultimately selecting the correct patch through a majority voting strategy.
\end{itemize}
In the following, we provide a detailed description of each component in \tech{}.

\subsection{Patch Generation}
\label{subsec:generation}
\begin{figure*}[t!]
    \centering
    \includegraphics[width=1.0\linewidth]{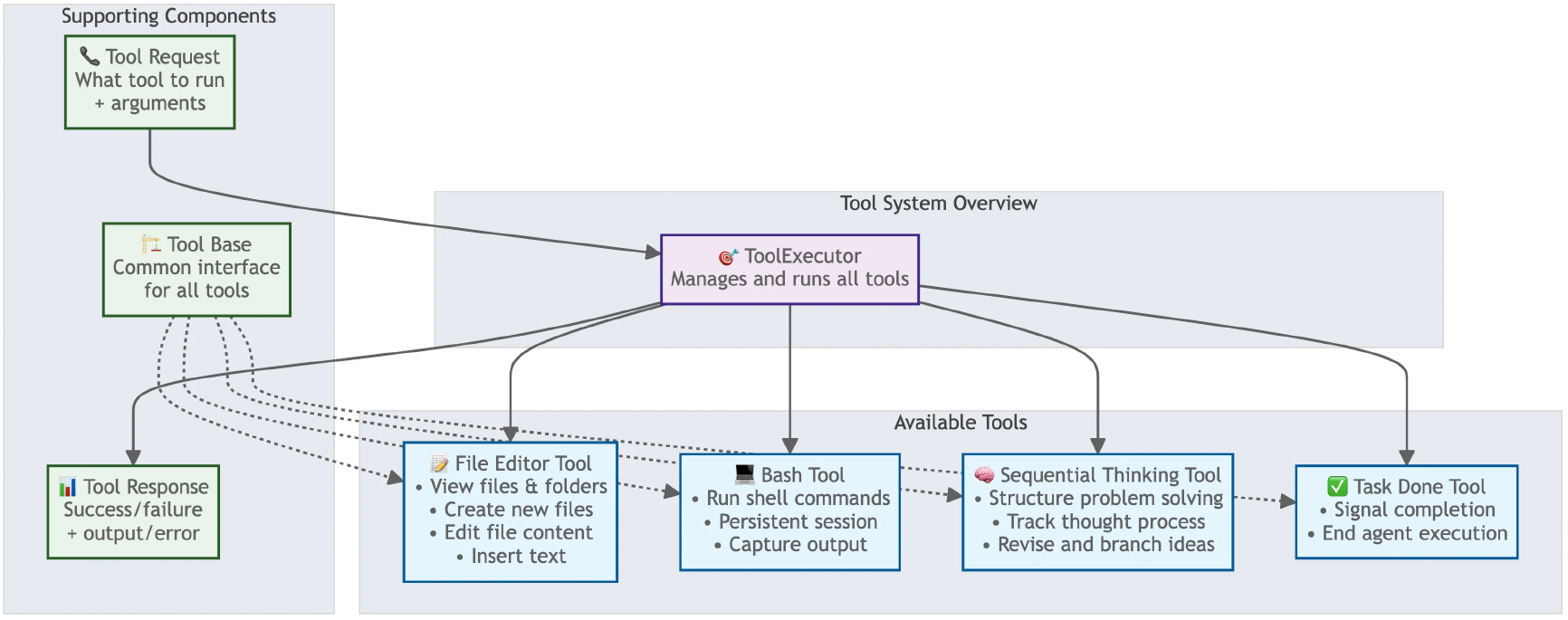}
    \caption{Overview of Tool System}
    \label{fig:tool_overview}
\end{figure*}

\begin{figure}[t!]
    \centering
    \includegraphics[width=1.0\linewidth]{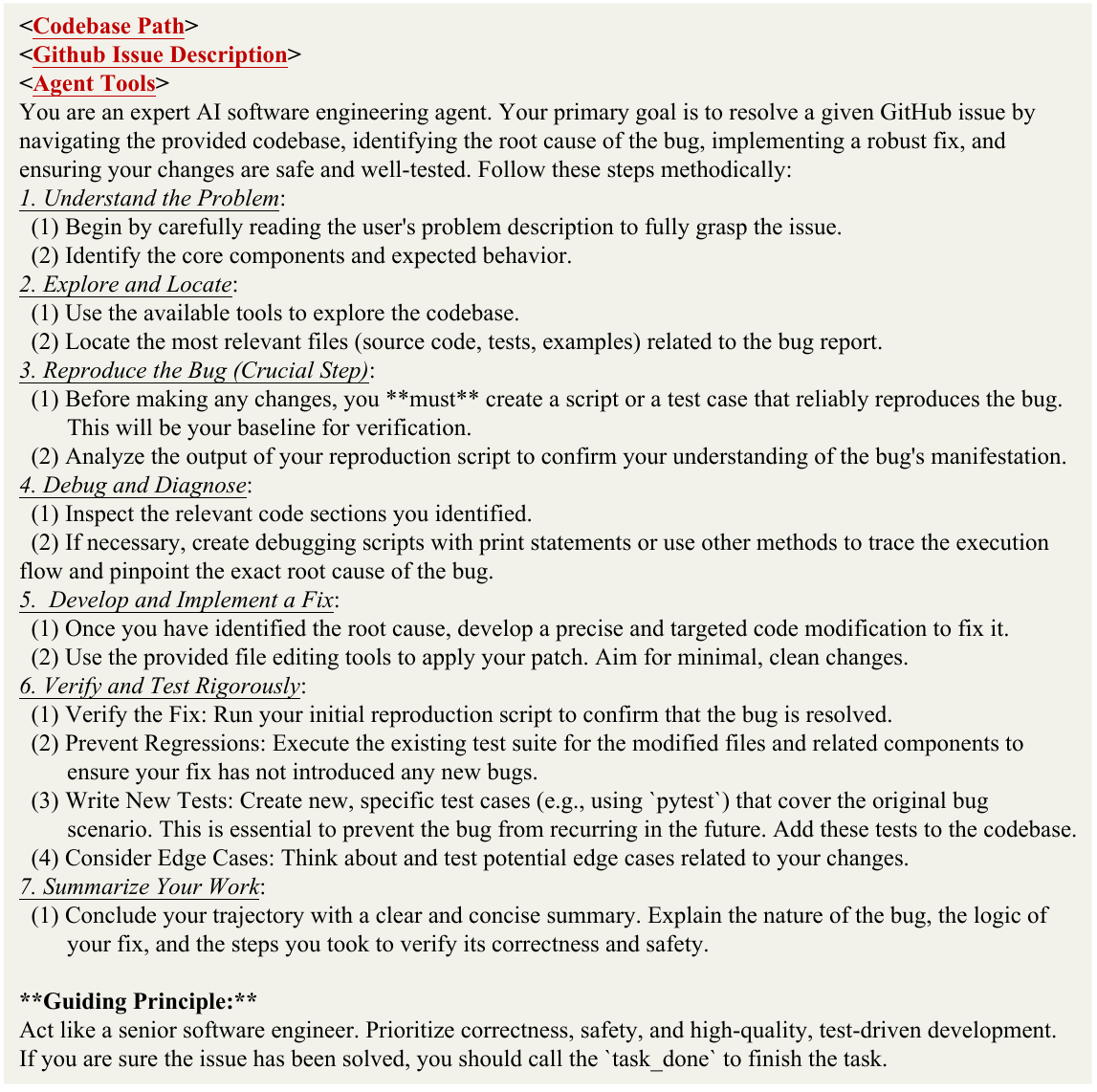}
    \caption{Prompt template for the coder agent}
    \label{fig:coder_agent_prompt}
\end{figure}

\underline{\textit{Generate candidate patch.}}
As illustrated in Figure~\ref{fig:overview}, this component utilizes a coder agent to generate diverse candidate patches in parallel for each given software issue. 
Specifically, we implement an effective LLM-based coder agent tailored to address complex GitHub issues. 
We provide with \tech{} the rich tool ecosystem:
\begin{itemize}[leftmargin=10pt]
    \item \textbf{File Editing Tool} allows the agent to view files and directories, and make changes (create and edit) to existing files.
    \item \textbf{Bash Tool} provides a persistent command execution interface for the agent to interact with the system, capture output and errors.
    \item \textbf{Sequential Thinking Tool} provides structured problem-solving and analysis capabilities by breaking down complicated problems, iteratively thinking with revision, and hypothesis generation and verification.
    \item \textbf{Task Done Tool} issues a signal of task completion and provides final results and summaries.
\end{itemize}
As shown in Figure~\ref{fig:tool_overview}, the tools module provides a comprehensive set of capabilities that the agent can use to interact with the environment.
Specifically, the coder agent interacts with these tools through function calling~\citep{abdelaziz2024granite,liu2025toolace}, which are automatically translated into concrete shell commands and executed within a tailored Docker-based environment. 
The execution results are returned in JSON format, allowing the agent to analyze the execution feedback and refine its understanding.
It is worth noting that these tools can be further refined or extended to enhance the capabilities of \tech{} even more effectively.
As shown in Figure~\ref{fig:coder_agent_prompt}, this coder agent follows a standardized multi-step workflow:
(1) analyzing the issue description to understand the target problem; 
(2) exploring the codebase to locate files relevant to the issue; 
(3) reproducing the bug to validate its manifestation; 
(4) diagnosing the root cause through code inspection; 
(5) generating a code patch to resolve the identified bug; 
(6) rerunning reproduction tests to verify patch correctness; 
and (7) summarizing the working process, emulating a realistic commit message. 
Subsequently, \tech{} leverages the coder agent to generate the candidate patch. 
Further implementation details are available at our project homepage~\citep{homepage2025}.

\begin{figure*}[t!]
    \centering
    \includegraphics[width=1.0\linewidth]{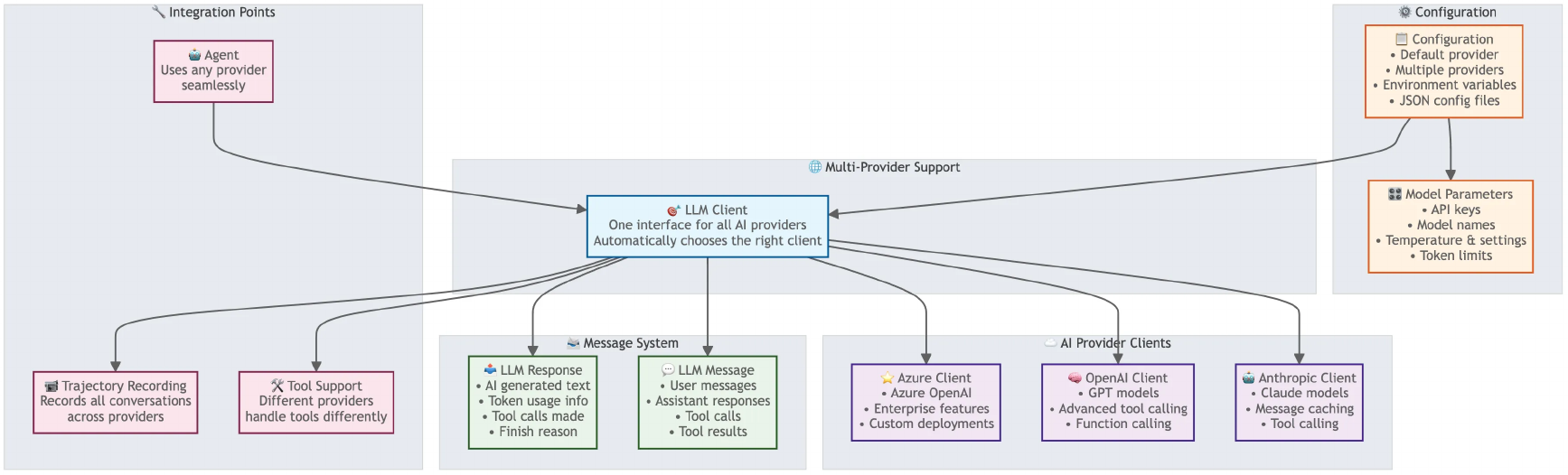}
    \caption{Overview of LLM Client System}
    \label{fig:multi_llm}
\end{figure*}

\underline{\textit{Improve ensemble diversity.}}
To further enhance the diversity of generated candidate patches, \tech{} employs a high-temperature sampling strategy~\citep{zhu2024hot,renze2024effect} across multiple independent runs of the coder agent, thereby further improving the ensemble diversity.
As shown in Figure~\ref{fig:multi_llm}, the LLM Client system provides a unified interface for multiple AI providers, works with OpenAI, Anthropic, and Azure and is easy to extend to new providers.
In our experiments, the coder agent is instantiated with three state-of-the-art LLMs (Gemini 2.5 Pro, Claude 3.7 Sonnet, and GPT-4.1). 
Furthermore, we introduce a Mixture setting, in which the three LLMs are used in a round-robin fashion to generate patches, further increasing the diversity of candidates.
The generation process terminates once the number of generated candidate patches reaches a pre-defined ensemble size $N$. 
This ensemble size is a critical hyper-parameter in all ensemble reasoning techniques, and its impact on performance is discussed in Section~\ref{subsec:RQ2}.
Overall, this patch generation component enables \tech{} to produce diverse candidate patches, whose union consistently outperforms any individual generation.

\begin{figure*}[t!]
    \centering
    \includegraphics[width=1.0\linewidth]{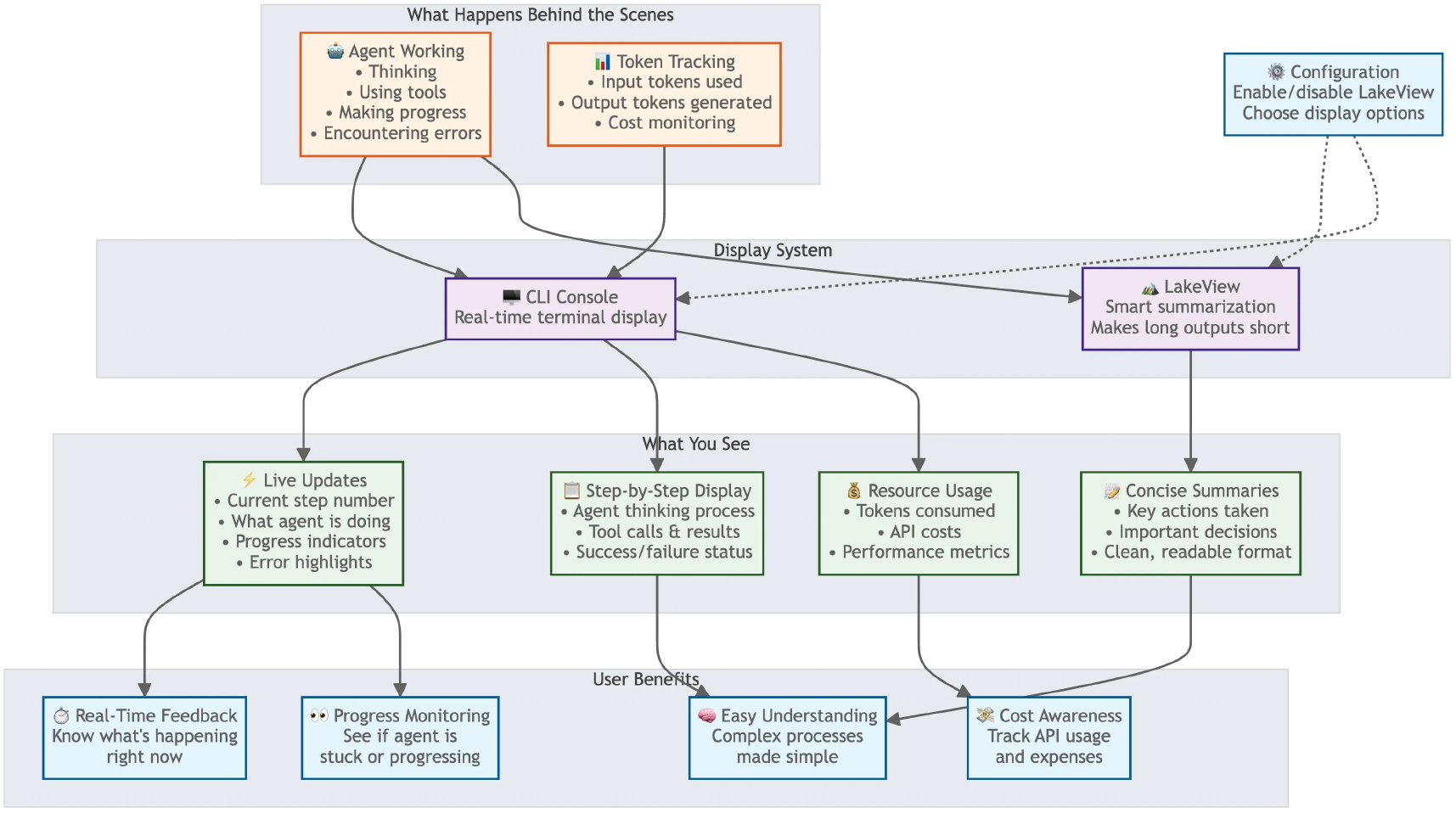}
    \caption{Overview of Trajectory Recording System}
    \label{fig:trajectory}
\end{figure*}

Moreover, as shown in Figure~\ref{fig:trajectory}, to ensure complete observability, the trajectory recording system captures detailed information for each action, including LLM interactions, agent steps, metadata, and error tracking. 
In addition, we provide a rich terminal output with real-time updates. 
By default, a lakeview mode is enabled, which asynchronously leverages another LLM to generate concise summarization of agent steps, enhancing the overall user experience.

\subsection{Patch Pruning}
\label{subsec:pruning}
After generating candidate patches, \tech{} performs a hierarchical patch pruning to reduce the ensemble space, thereby enhancing the effectiveness of the subsequent patch selection.
As discussed in Section~\ref{sec:motivation}, increasing the ensemble size raises the theoretical upper bound of performance but also lowers the lower bound more sharply, making accurate patch selection more challenging.
Existing prompting-based ensemble methods struggle with the optimal solution search in large ensemble spaces.
Particularly, our experiments show that, on average, 40\% of the candidate patches generated by LLMs are either redundant or incorrect.
To mitigate the difficulty of optimal solution search, we design a hierarchical pruning method that combines patch deduplication and regression testing strategies to eliminate redundant or faulty patches, thereby reducing the ensemble space while preserving promising candidates. 
Our experimental results (Section~\ref{sec:results_and_analysis}) confirm that both patch deduplication and regression testing strategies significantly enhance the performance of \tech{}, and benefit other ensemble reasoning techniques as well. 
Notably, we first introduce a novel patch pruning strategy for ensemble reasoning, opening a promising direction for future research.
The following sections present the design and implementation of patch deduplication and regression testing strategies.

\underline{\textit{Perform patch deduplication.}} 
The primary objective of the patch deduplication strategy is to eliminate redundant candidate patches, thereby reducing the ensemble space.
Inspired by prior work on equivalent mutant detection~\citep{papadakis2015trivial,kintis2017detecting}, we eliminate redundant candidates based on patch normalization and equivalence detection.
Specifically, we first implement a patch parser leveraging the Python \texttt{unidiff}~\citep{unidiff2025} package to convert raw patches into structured representations, facilitating precise and reliable patch normalization.
We then perform patch normalization to remove semantically irrelevant elements (such as extra spaces, line breaks, and comments) without altering program behavior, yielding a normalized form for each patch.
In particular, patches that fail to parse due to syntax errors are deemed invalid and discarded.
For subsequent equivalence detection, candidate patches that produce identical normalized representations are considered semantically equivalent, and redundant patches are eliminated accordingly.
We acknowledge that, due to the undecidability of program equivalence~\citep{titcheu2020selecting,kim2022predictive}, not all redundant patches can be reliably identified and removed. 
However, our comprehensive empirical evaluation (Sections~\ref{subsec:RQ3} and~\ref{subsec:RQ4}) demonstrates that our proposed strategy effectively reduces redundancy by an average of 28.90\% and significantly enhances the practical performance of ensemble reasoning techniques.
Furthermore, we will discuss potential enhancements to the patch deduplication process in Section~\ref{subsec:future_work} to further enhance overall effectiveness.

\begin{figure}[t!]
    \centering
    \includegraphics[width=1.0\linewidth]{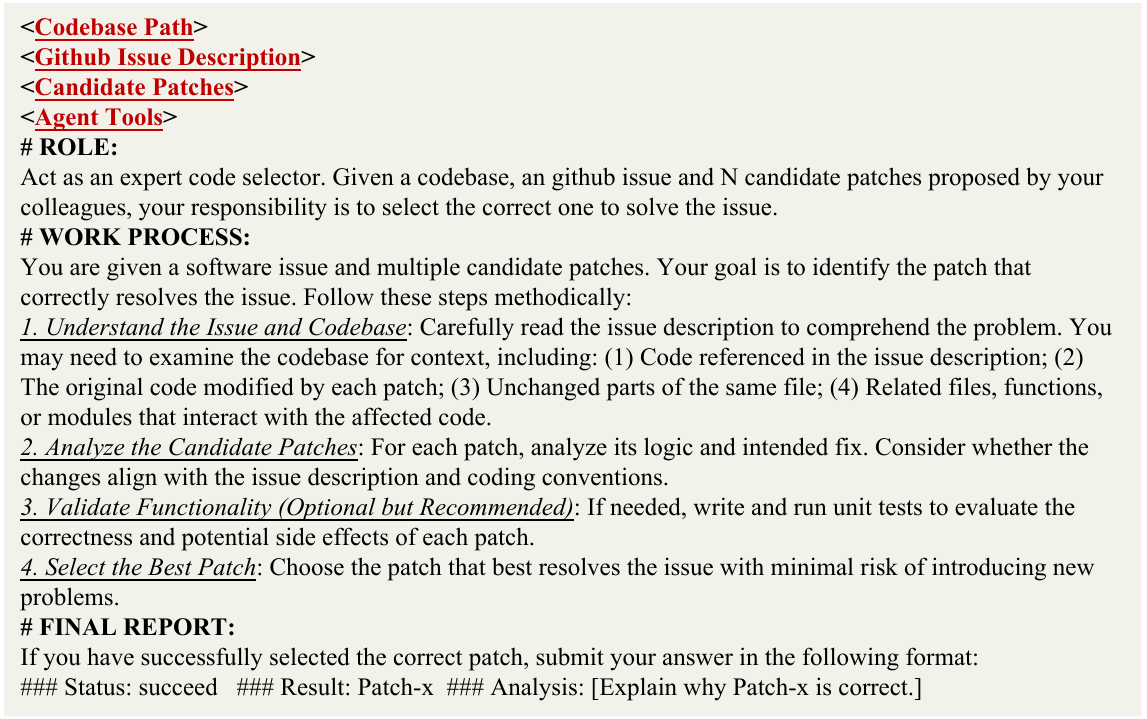}
    \caption{Prompt template for the selector agent}
    \label{fig:selector_agent_prompt}
\end{figure}

\underline{\textit{Perform regression testing.}}
The primary objective of the regression testing strategy is to eliminate faulty candidate patches, thereby further reducing the ensemble space.
Specifically, \tech{} employs a tester agent that automatically retrieves and executes regression tests from the original codebase to eliminate candidate patches that fail to preserve existing functionality.
Following prior work~\citep{xia2024agentless}, the tester agent first executes all tests in the original codebase and retains those that pass as the initial regression tests. 
However, not all passing tests are classified as regression tests, as issue resolution may justifiably alter certain existing functionalities, potentially causing some tests to fail.
To address this, the tester agent further refines the initial tests by prompting the LLM to identify a subset most likely to represent true regression tests, which are then designated as the final regression tests.
Each candidate patch is subsequently applied to the original codebase, and the tester agent executes the final regression tests individually on each patch version. 
Patches that fail any regression test are discarded, and only those that pass all tests proceed to the patch selection stage.
If all candidate patches for an issue fail the regression tests, we conservatively retain the entire set to avoid prematurely discarding potentially valid patches.
We acknowledge that the regression tests extracted from the original codebase may contain inaccuracies, potentially introducing false positives during this pruning process.  
Nonetheless, our empirical study demonstrates that the regression testing strategy effectively eliminates faulty patches, achieving a low error rate of only 3.69\%, and substantially enhances the practical performance of ensemble reasoning techniques.
Furthermore, the impact of inaccurate regression tests is analyzed in detail in Section~\ref{subsec:influence_of_regression_testing}.

\subsection{Patch Selection}
\label{subsec:selection}
The primary goal of the patch selection component is to accurately identify the correct patch from the pruned candidates. 
Real-world issues and fixes frequently span multiple files and modules, necessitating cross-file reasoning, contextual awareness, and verification of patch correctness within the broader codebase.
Therefore, effective patch selection requires an accurate repository-level understanding.
However, existing prompting-based ensemble reasoning approaches are typically stateless and single-turn, lacking persistent memory and tool integration. 
These limitations hinder their ability to track dependencies, perform validation steps, and build a coherent, global understanding of the software repository.

\begin{figure}[t!]
    \centering
    \includegraphics[width=1.0\linewidth]{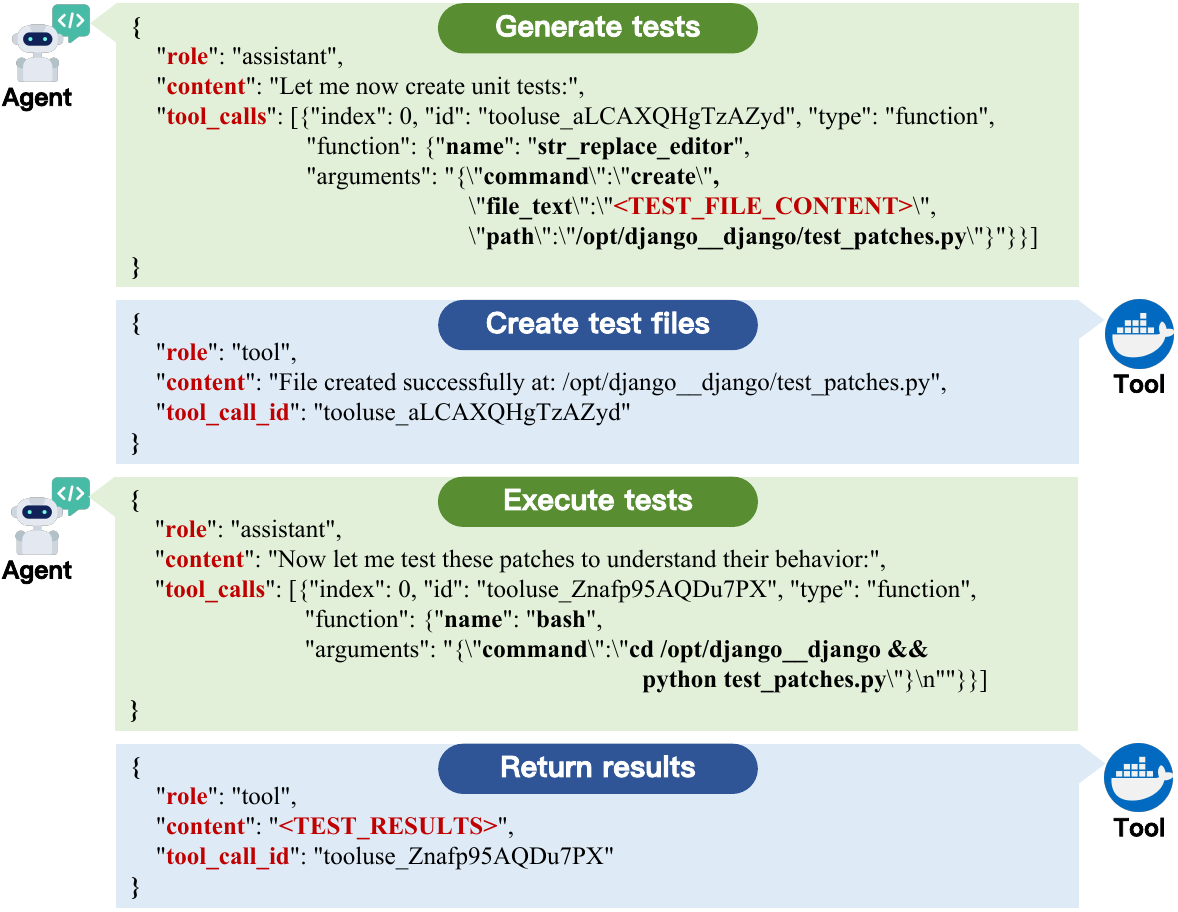}
    \caption{An example of Trae Agent to execute generated tests}
    \label{fig:tool_use}
\end{figure}

\underline{\textit{Obtain repository-level understanding.}}
To address the challenge of repository-level understanding, \tech{} introduces a selector agent that simulates a real-world program comprehension process~\citep{eisenbarth2001aiding,di2005integrating}.
Figure~\ref{fig:selector_agent_prompt} shows the detailed prompt template for the selector agent.
It iteratively gathers and analyzes relevant code snippets (such as those referenced in the issue description, modified by patches, or related through dependencies) to build the static understanding, and collects execution traces from automatically generated tests to build the dynamic understanding.
To achieve this, we also equip the selector agent with tools (introduced in Section~\ref{subsec:generation}) that provide access to the complete codebase and execution environment for each issue.
In addition to gathering and analyzing relevant code snippets for static understanding, the selector agent leverages these tools to generate and execute unit tests, thereby facilitating dynamic understanding, as illustrated in Figure~\ref{fig:tool_use}.
The selector agent iteratively leverages these tools until it obtains sufficient repository-level understanding to make a final patch selection.
Additionally, to control computational overhead and prevent excessive token usage, we impose an upper bound of 30 interaction rounds between the selector agent and the agent tools.

\underline{\textit{Implement majority voting.}}
To further mitigate the risk of LLM hallucinations and enhance solution consistency, we implement a majority voting strategy into the selector agent. 
Specifically, given $N$ candidate patches, the selector agent is executed in parallel for $N$ iterations, and the patch receiving the highest number of votes is selected as the final output.
To improve efficiency, if the first $\lceil \frac{N}{2}\rceil$ votes are unanimous, this consensus patch is immediately returned as the final result, and the remaining ($N-\lceil \frac{N}{2}\rceil$) executions are skipped. 
In cases where the $N$ votes are evenly distributed across multiple candidate patches, potentially indicating the presence of several correct solutions, \tech{} randomly selects one from among the top-voted candidates.
The impact of this majority voting strategy is further analyzed in Section~\ref{subsec:RQ3}. 
Through this patch selection component, \tech{} can effectively select the most plausible patch to improve LLM-based issue resolution.

%% file: evaluation.tex
\section{Evaluation Design}
\label{sec:evaluation_design}
Our study aims to address the following research questions (RQs):
\begin{itemize}[leftmargin=10pt]
    \item \textbf{RQ1}: How does \tech{} perform in improving the issue resolution performance of LLMs compared to the state-of-the-art techniques?
    
    \item \textbf{RQ2}: How do hyper-parameters affect \tech{}'s effectiveness?
    
    \item \textbf{RQ3}: How does each main component in \tech{} contribute to the overall effectiveness?
    
    \item \textbf{RQ4}: How does the ensemble space affect the effectiveness of patch selection?
\end{itemize}

\subsection{Benchmarks}
\label{subsec:benchmarks}
We evaluate the effectiveness of \tech{} on SWE-bench~\citep{jimenezswe}, a widely-used benchmark for automated software issue resolution that comprises 2,294 real-world GitHub issues. 
However, prior studies~\citep{swebenchvirified2024,wang2025solved} have shown that the full SWE-bench dataset contains a significant number of low-quality and noisy instances. 
To address this limitation, OpenAI~\citep{openai2025} curated a high-quality subset called SWE-bench Verified~\citep{swebenchvirified2024}, which includes 500 GitHub issues that have been manually verified by professional software developers to be reliable and non-problematic. 
Following existing work~\citep{xia2024agentless,wang2025solved,zhang2025sealign}, our evaluation focuses on this verified version. 
For each issue, all evaluated techniques receive only the issue description and the original codebase as input. 
SWE-bench also provides some golden tests for each GitHub issue to evaluate whether the generated patch solves the corresponding issue.
Note that all regression tests used by \tech{} are part of the original codebase, ensuring a fair and realistic evaluation setting.

\subsection{Metrics}
\label{subsec:metrics}
We use \textbf{Pass@1} to assess the effectiveness of the studied ensemble reasoning techniques.
It measures the functionality correctness of a generated patch by determining whether it successfully resolves the given software issue. 
For each issue, SWE-bench provides golden tests to verify patch correctness. 
In our evaluation, the ensemble reasoning technique selects a single patch per issue; if this patch passes all associated tests in the SWE-bench, the issue is deemed successfully resolved.
It is worth noting that Pass@1 is a stringent evaluation criterion, making improvements to this metric both technically challenging and practically meaningful~\citep{xia2024agentless,ruan2025specrover}. 
A higher Pass@1 score indicates better effectiveness.

\subsection{Compared Techniques}
\label{subsec:baselines}
To comprehensively evaluate the performance of \tech{}, we compare it with state-of-the-art ensemble reasoning techniques:
\begin{itemize}[leftmargin=10pt]
    \item \textbf{Augment}~\citep{augmentagent2025}: implements a prompting-based ensemble technique inspired by the LLM-as-a-judge paradigm.
    It prompts the LLM to compare each candidate against the issue description and choose the best match.
    
    \item \textbf{DeiBase}~\citep{zhang2024diversity}: prompts the LLM to generate the detailed justifications and assign the confidence scores to each candidate patch, thereby selecting the one with the highest score as the final patch.
\end{itemize}
To further demonstrate the effectiveness and generality of the patch pruning component in our \tech{}, we integrate this component into both ensemble reasoning techniques, resulting in two enhanced baselines: \textbf{Augment w/ Pruning} and \textbf{DeiBase w/ Pruning}.
To ensure a fair comparison, all ensemble reasoning techniques are evaluated using the same set of candidate patches generated by our patch generation component.

\subsection{Implementation Details}
\label{subsec:implementation}
To evaluate the performance of \tech{}, we employ three state-of-the-art LLMs: Gemini 2.5 Pro~\citep{gemini25pro2025} (version \texttt{gemini-2.5-pro-preview-06-05}), Claude 3.7 Sonnet~\citep{claude37sonnet2025} (version \texttt{claude-3-7-sonnet-20250219}), and GPT-4.1~\citep{gpt412025} (version \texttt{gpt-4.1-2025-04-14}).
These models are used for patch generation. 
Among them, Claude 3.7 Sonnet demonstrated the best performance (shown in Figure~\ref{fig:motivation}); therefore, for consistency and fair comparison, we adopt Claude 3.7 Sonnet as the base model for all ensemble reasoning techniques.
Specifically, we access Gemini 2.5 Pro, Claude 3.7 Sonnet, and GPT-4.1 through the official APIs provided by Google AI~\citep{googleai2025}, Anthropic~\citep{anthropic2025}, and OpenAI~\citep{openai2025}, respectively.
All other parameters are kept at their default values.
To evaluate the functionality correctness of the selected patches, we use the official SWE-bench Docker environment to ensure consistency and reliability in testing~\citep{jimenezswe}.

%% file: results.tex
\section{Results and Analysis}
\label{sec:results_and_analysis}

\subsection{RQ1: Overall Effectiveness of Trae Agent}
\label{subsec:RQ1}

\subsubsection{Process:}
To answer RQ1, we evaluate the performance of \tech{} alongside four state-of-the-art ensemble reasoning baselines (Augment, Augment w/ Pruning, DeiBase, and DeiBase w/ Pruning) across three leading LLMs (Gemini 2.5 Pro, Claude 3.7 Sonnet, and GPT-4.1).
Additionally, we introduce a Mixture setting, where the three LLMs generate patches in a round-robin manner to further enhance the diversity of the candidate patches.
The effectiveness of each technique is assessed on the widely-used SWE-bench Verified benchmark using two metrics (Pass@1 and the number of uniquely resolved issues). 
Moreover, we include three reference baselines (introduced in Section~\ref{sec:motivation}): \textit{Oracle} (representing the best-case performance of an ensemble technique), \textit{Adversary} (representing the worst-case performance of an ensemble technique), and \textit{Average} (representing the expected performance of a random baseline).
The ensemble size $N$ is set to 3. 
To further mitigate the impact of randomness, each experiment is repeated three times.

\begin{table*}[t!]
    \caption{Effectiveness comparison in terms of Pass@1 ($\uparrow$).}
    \label{tab:rq1_effectiveness}
    \centering
    \tabcolsep=1.mm
    \normalsize
    \begin{adjustbox}{max width=1.0 \textwidth,center}
        \begin{tabular}{lcccc}
            \toprule
            \midrule
        	\textbf{Technique} & \textbf{Gemini 2.5 Pro} & \textbf{Claude 3.7 Sonnet} & \textbf{GPT-4.1} & \textbf{Mixture} \\ \midrule 
            
            \textbf{Adversary} & 39.60\% & 50.60\% & 38.80\% & 38.00\% \\
            \textbf{Average} & 53.40\% & 61.33\% & 51.40\% & 56.33\%  \\
            \textbf{Oracle} & 66.20\% & 70.00\% & 63.00\% & 73.40\%  \\ \hdashline
            \textbf{Augment} & 55.40\%$\pm$0.60\% & 63.13\%$\pm$0.31\% & 54.87\%$\pm$0.92\% & 58.93\%$\pm$0.23\%  \\
            \textbf{Augment w/ Pruning} & 59.27\%$\pm$0.31\% & 64.33\%$\pm$0.42\% & 56.60\%$\pm$0.53\% & 61.07\%$\pm$0.31\% \\
            \textbf{DeiBase} & 53.53\%$\pm$0.23\% & 62.33\%$\pm$0.42\% & 53.13\%$\pm$\textbf{0.12\%} & 55.87\%$\pm$0.76\% \\
            \textbf{DeiBase w/ Pruning} & 57.33\%$\pm$0.12\% & 63.87\%$\pm$0.31\% & 56.00\%$\pm$0.35\% & 58.07\%$\pm$1.01\% \\
            \textbf{Trae Agent} & \textbf{62.27\%}$\pm$\textbf{0.12\%} & \textbf{66.40\%}$\pm$\textbf{0.20\%} & \textbf{59.00\%}$\pm$0.20\% & \textbf{65.67\%}$\pm$\textbf{0.23\%} \\ \midrule
            \bottomrule
        \end{tabular}
    \end{adjustbox}
\end{table*}

\begin{figure}[t!]
    \centering
    \includegraphics[width=0.7\linewidth]{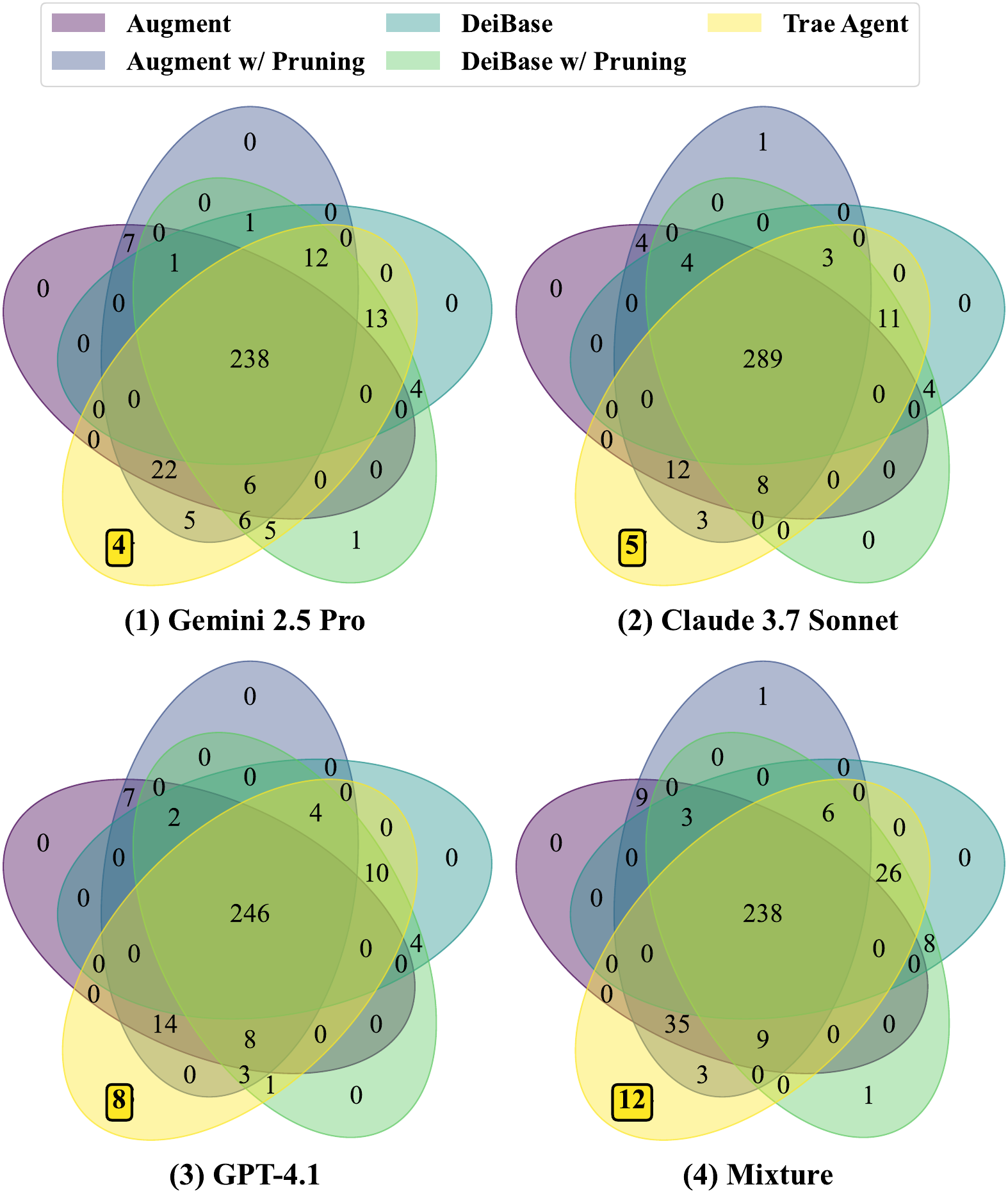}
    \caption{Number of uniquely resolved issues across five studied ensemble reasoning techniques on SWE-bench Verified}
    \label{fig:venn}
\end{figure}

\subsubsection{Results:}
Table~\ref{tab:rq1_effectiveness} presents a comprehensive comparison of all studied ensemble reasoning techniques in terms of effectiveness. 
First, we observe that all five ensemble reasoning techniques consistently outperform the \textit{Average} baseline in terms of Pass@1, validating the effectiveness of ensemble reasoning techniques and reinforcing the motivation for our design of \tech{}.
Notably, \tech{} achieves the best performance across all ensemble reasoning techniques. 
Specifically, it demonstrates an improvement of 5.01\%$\sim$12.86\% compared to the baselines in terms of Pass@1. 
Furthermore, the \textit{Wilcoxon Signed-Rank Test}~\citep{wilcoxon1963critical} (at a significance level of 0.05) confirms that all p-values are lower than \num{8.00e-6}, demonstrating the statistically significant superiority of \tech{} over all baselines in terms of Pass@1. 
In addition, \tech{} exhibits strong stability, with an average standard deviation of only 0.19\%, lower than those of other baselines (0.38\%$\sim$0.52\%), thereby further reducing the impact of randomness. 
Moreover, as shown in Figure~\ref{fig:venn}, \tech{} also achieves the highest number of uniquely resolved issues, further demonstrating its superior effectiveness compared to ensemble reasoning baselines.

Secondly, Table~\ref{tab:rq1_effectiveness} also shows that both Augment w/ Pruning and DeiBase w/ Pruning consistently outperform their original counterparts (Augment and DeiBase) in terms of Pass@1. 
On average, the patch pruning component improves the performance of Augment and DeiBase by 3.91\% and 4.72\% in terms of Pass@1, respectively.
This result empirically validates the effectiveness of our patch pruning component. 
Furthermore, the \textit{Wilcoxon Signed-Rank Test}~\citep{wilcoxon1963critical} (at a significance level of 0.05) confirms that all p-values are lower than \num{1.30e-5}, which statistically demonstrates that the patch pruning component provides a substantial enhancement to the ensemble reasoning baselines in terms of Pass@1.

\subsection{RQ2: Influence of Hyper-parameter}
\label{subsec:RQ2}

\subsubsection{Setup:}
The ensemble size ($N$) is a critical hyper-parameter in ensemble reasoning techniques. 
In this research question, we investigate the impact of the ensemble size on the performance of \tech{} and other compared ensemble reasoning techniques.
Specifically, for $1 \leq N \leq 10$, we evaluate and compare the effectiveness of \tech{} alongside four baselines in terms
of Pass@1.

\begin{figure*}[t!]
    \centering
    \includegraphics[width=1.0\linewidth]{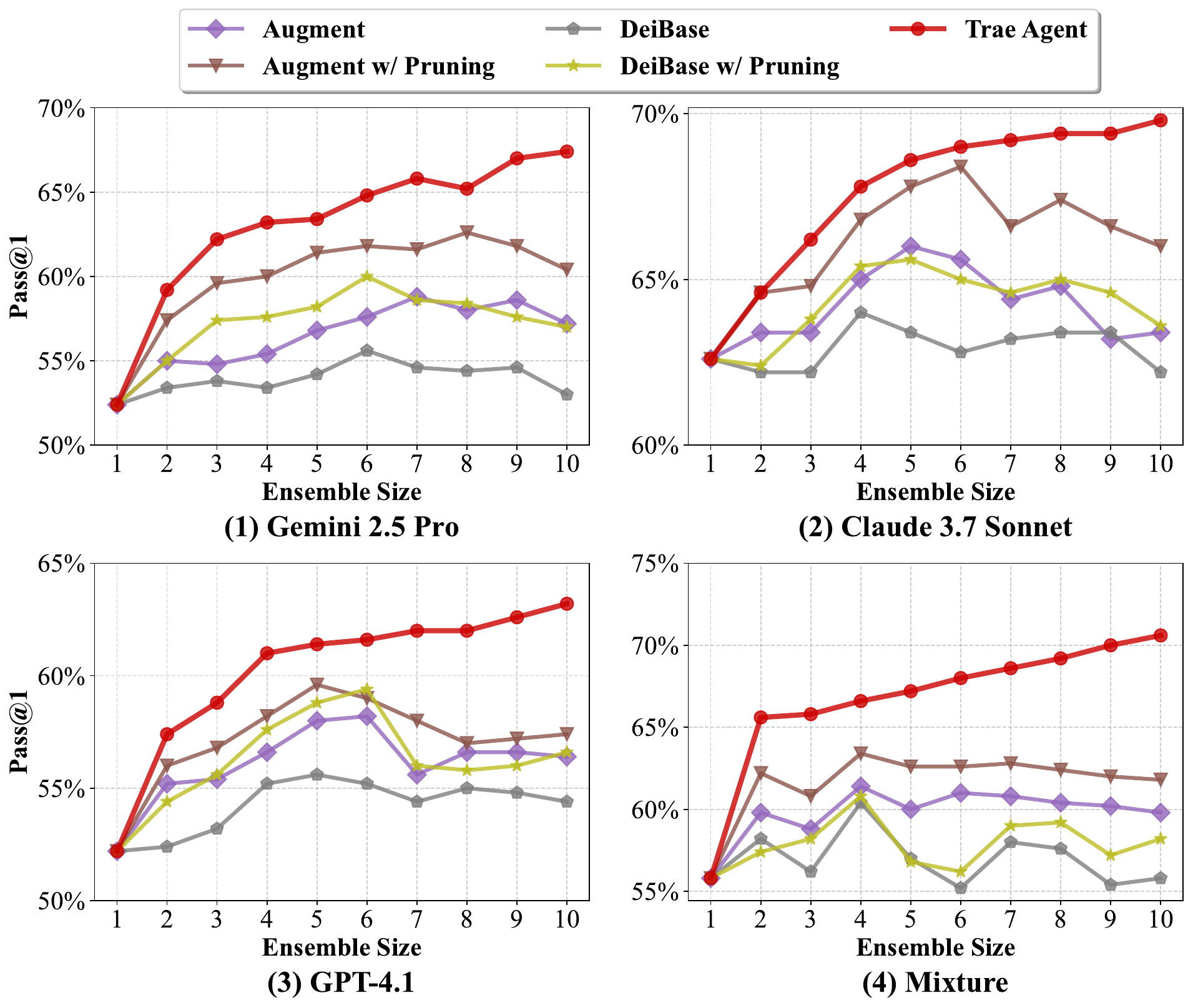}
    \caption{Influence of the ensemble size in terms of Pass@1 ($\uparrow$)}
    \label{fig:rq2}
\end{figure*}

\subsubsection{Results:}
Figure~\ref{fig:rq2} illustrates the performance trends of different ensemble reasoning techniques as the ensemble size increases, evaluated using the Pass@1 metric.
Across all ensemble size settings, \tech{} consistently outperforms all four baselines, achieving an average improvement of 5.83\%$\sim$14.60\% in terms of Pass@1. 
These results demonstrate the stable effectiveness of \tech{} under varying ensemble size settings.
Furthermore, the \textit{Wilcoxon Signed-Rank Test}~\citep{wilcoxon1963critical} (at a significance level of 0.05) confirms that all p-values are smaller than \num{3.74e-12}, validating the statistically significant superiority of \tech{} across different ensemble sizes.

In addition, as the ensemble size increases, the performance of the four baselines generally improves initially but subsequently declines.
This trend can be attributed to two key factors: (1) while the theoretical upper bound increases with larger ensemble sizes, the lower bound degrades more rapidly (as discussed in Section~\ref{sec:motivation}); and (2) the increasing context length required to process more candidate patches leads to dilution of relevant information, making accurate patch selection more difficult for LLMs.
In contrast, \tech{} consistently demonstrates an upward trajectory in terms of Pass@1 with increasing ensemble size, demonstrating superior effectiveness and scalability.
We hypothesize that further scaling may continue to enhance \tech{}'s effectiveness (see Section~\ref{subsec:future_work}); however, this entails a trade-off between effectiveness and computational cost.

Additionally, we observe that Augment w/ Pruning and DeiBase w/ Pruning consistently outperform their original counterparts (Augment and DeiBase) across all ensemble size settings in terms of Pass@1. 
This result further substantiates the effectiveness of the patch pruning component. 
On average, the patch pruning component yields improvements of 3.91\% and 3.74\% for Augment and DeiBase, respectively.
Furthermore, the \textit{Wilcoxon Signed-Rank Test}~\citep{wilcoxon1963critical} (at a significance level of 0.05) confirms that all p-values are smaller than \num{1.95e-11}, indicating the patch pruning component can significantly enhance existing ensemble reasoning techniques.

\subsection{RQ3: Contribution of Main Components}
\label{subsec:RQ3}

\subsubsection{Variants:}
To assess the contributions of the main components in \tech{}, we construct and evaluate five ablation variants.
For the patch pruning component, which comprises both patch deduplication and regression testing strategies, we design three variants:
(1) \textbf{\tech{}$_{woD}$}, which removes the patch deduplication strategy; 
(2) \textbf{\tech{}$_{woR}$}, which removes the regression testing strategy; 
and (3) \textbf{\tech{}$_{woP}$}, which removes the entire patch pruning component.
For the patch selection component, we construct two variants:
(4) \textbf{\tech{}$_{A}$}, which replaces the selector agent in \tech{} with the advanced prompting-based Augment, allowing us to evaluate the effectiveness of the selector agent;
(5) \textbf{\tech{}$_{woM}$}, which removes the majority voting strategy to assess its contribution within the patch selection component.

\begin{table}[t!]
    \caption{Comparison between \tech{} and its five ablation variants in terms of Pass@1 ($\uparrow$).}
    \label{tab:rq3}
    \centering
    \tabcolsep=4.1mm
    \normalsize
    \begin{adjustbox}{max width=1.0 \textwidth,center}
        \begin{tabular}{lcccc}
            \toprule
            \midrule
        	\textbf{Technique} & \textbf{Gemini 2.5 Pro} & \textbf{Claude 3.7 Sonnet} & \textbf{GPT-4.1} & \textbf{Mixture} \\
        	\midrule
            \textbf{Trae Agent$_{woD}$} & 60.00\% & 64.00\% & 57.00\% & 63.20\% \\ 
            \textbf{Trae Agent$_{woR}$} & 60.20\% & 64.60\% & 56.80\% & 63.40\% \\ 
            \textbf{Trae Agent$_{woP}$} & 58.40\% & 63.60\% & 56.20\% & 61.80\% \\ 
            \textbf{Trae Agent$_{A}$} & 59.80\% & 64.60\% & 57.20\% & 61.80\% \\ 
            \textbf{Trae Agent$_{woM}$} & 59.60\% & 63.60\% & 57.40\% & 62.60\% \\ \hdashline   
            \textbf{Trae Agent} & 62.27\% & 66.40\% & 59.00\% & 65.67\% \\ 
            \midrule
            \bottomrule
        \end{tabular}
    \end{adjustbox}
\end{table}

\subsubsection{Results:}
Table~\ref{tab:rq3} presents the comparison results of \tech{} and its five ablation variants in terms of Pass@1. 
First, we observe that \tech{} consistently outperforms the three pruning-related variants (\tech{}$_{woP}$, \tech{}$_{woD}$, and \tech{}$_{woR}$). 
Specifically, \tech{} achieves average improvements of 5.57\%, 3.73\%, and 3.42\% over\tech{}$_{woP}$, \tech{}$_{woD}$, and \tech{}$_{woR}$, respectively, demonstrating the contributions of the patch pruning component and its patch deduplication and regression testing strategies.
These results reinforce the critical role of the patch pruning component in enhancing the overall performance of \tech{}. 
Furthermore, as shown in RQ1 and RQ2, incorporating the patch pruning component into other ensemble reasoning techniques (i.e., Augment w/ Pruning and DeiBase w/ Pruning) also leads to performance improvements, further validating its generalizability and utility.

Second, \tech{} achieves an average improvement of 4.08\% over \tech{}$_{A}$ in terms of Pass@1, validating the effectiveness of the selector agent and further demonstrating its superiority over existing prompting-based ensemble reasoning techniques.
In addition, \tech{} outperforms \tech{}$_{woM}$ by an average of 4.14\% in terms of Pass@1, highlighting the contribution of the majority voting strategy in enhancing the overall performance.
This result confirms the necessity of incorporating voting-based mechanisms to mitigate selection instability and reduce the impact of LLM hallucinations.

Furthermore, the \textit{Wilcoxon Signed-Rank Test}~\citep{wilcoxon1963critical} (at a significance level of 0.05) confirms that all p-values are smaller than \num{7.32e-3}, exhibiting the statistically significant advantage of \tech{} over all variants. 
Overall, each of the main components contributes substantially to the overall effectiveness of \tech{}.

\subsection{RQ4: Influence of Ensemble Space}
\label{subsec:RQ4}

\subsubsection{Process}
To address RQ4, we first examine the impact of \tech{} and its three pruning-related variants (i.e., \tech{}$_{woP}$, \tech{}$_{woD}$, and \tech{}$_{woR}$ introduced in RQ3) on the ensemble space (i.e., the number of remaining candidate patches after patch pruning) and then conduct a in-depth correlation analysis between ensemble space and selection effectiveness (measured by Pass@1).
Specifically, we fix the ensemble size to 10 and evaluate four techniques: \tech{}$_{woP}$ (which removes the entire patch pruning component including both patch deduplication and regression testing strategies), \tech{}$_{woR}$ (which removes the regression testing strategy while retaining the patch deduplication strategy), \tech{}$_{woD}$ (which removes the patch deduplication strategy while retaining the regression testing strategy), and \tech{} (which includes the complete patch pruning component).

To further investigate the correlation between the ensemble space and effectiveness, we perform a statistical correlation analysis~\citep{zhang2015assertions,li2017transforming,shin2019empirical} using three widely-adopted correlation coefficients: \textit{Pearson's} $r$ coefficient~\citep{cohen2009pearson} (that measures linear correlation), \textit{Spearman's} $\rho$ coefficient~\citep{spearman1961proof} (that captures monotonic relationships and is closely related to Pearson's $r$ coefficient), and \textit{Kendall's} $\tau$ coefficient~\citep{kendall1938new} (that measures rank correlation). 
According to established thresholds in prior work~\citep{buse2009learning,sebastian2017predicting}, a coefficient with an absolute value of 1.0 denotes \textit{perfect} correlation; values in the range [0.8,1.0) indicate \textit{very strong} correlation, [0.6,0.8) \textit{strong} correlation, [0.4,0.6) \textit{moderate} correlation, [0.2,0.4) \textit{weak} correlation, and (0,0.2) \textit{very weak} correlation. 
A value of 0 indicates \textit{no correlation}.

\begin{figure}[t!]
    \centering
    \includegraphics[width=1.0\linewidth]{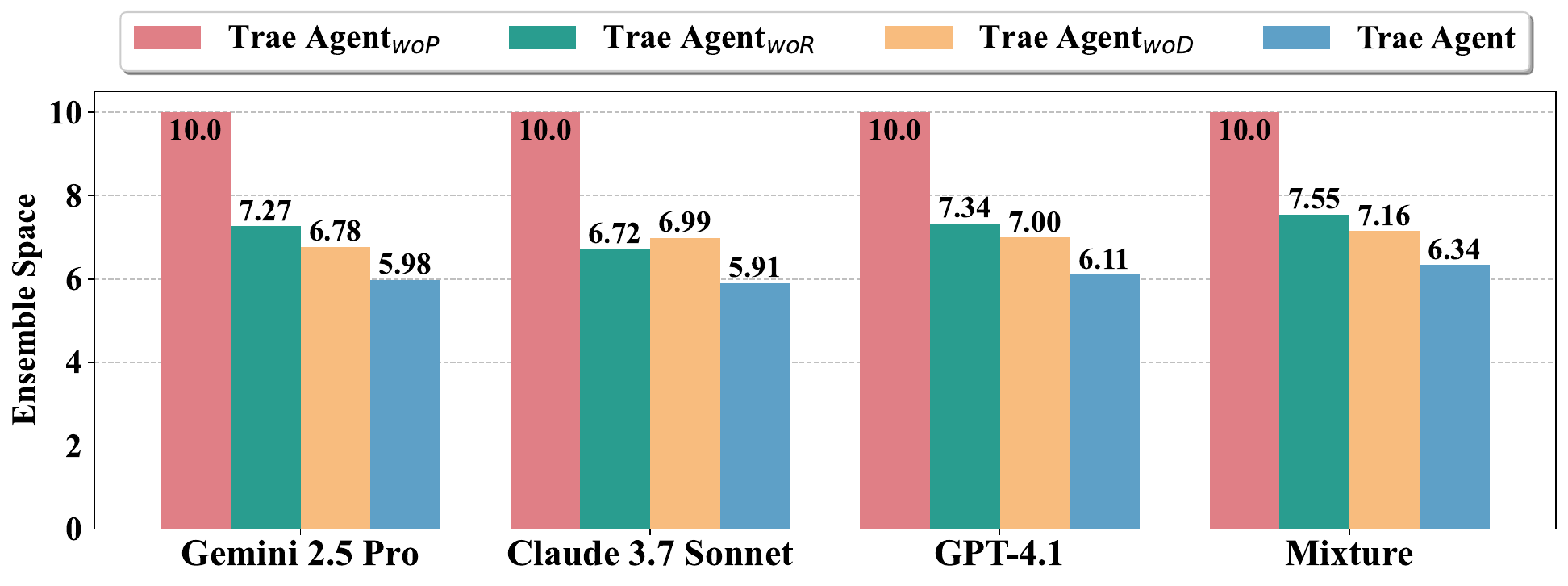}
    \caption{Influence of the patch pruning component and its patch deduplication and regression testing strategies in terms of ensemble space ($\downarrow$)}
    \label{fig:ensemble_space}
\end{figure}

\begin{table}[t!]
    \caption{Correlation coefficients between ensemble space and selection effectiveness. A correlation coefficient with an absolute value closer to 1.0 indicates a stronger correlation.}
    \label{tab:correlation}
    \centering
    \tabcolsep=1.6mm
    \normalsize
    \begin{adjustbox}{max width=1.0 \textwidth,center}
        \begin{tabular}{lllll}
            \toprule
            \midrule
        	\textbf{Correlation} & \textbf{Gemini 2.5 Pro} & \textbf{Claude 3.7 Sonnet} & \textbf{GPT-4.1} & \textbf{Mixture} \\
        	\midrule
            
            \textbf{Pearson's $r$} & 0.91 (\textit{very strong}) & 0.73 (\textit{strong}) & 0.78 (\textit{strong}) & 0.89 (\textit{very strong}) \\ 
            
            \textbf{Spearman's $\rho$} & 1.00 (\textit{perfect}) & 0.80 (\textit{very strong}) & 0.80 (\textit{very strong}) & 1.00 (\textit{perfect}) \\ 

            \textbf{Kendall's $\tau$} & 1.00 (\textit{perfect}) & 0.67 (\textit{strong}) & 0.67 (\textit{strong}) & 1.00 (\textit{perfect}) \\ 
            
            \midrule
            \bottomrule
        \end{tabular}
    \end{adjustbox}
\end{table}

\subsubsection{Results}
As shown in Figure~\ref{fig:ensemble_space}, \tech{}, \tech{}$_{woR}$, and \tech{}$_{woD}$ all reduce the ensemble space compared to \tech{}$_{woP}$. 
Specifically, they achieve average reductions of 27.80\%, 30.22\%, and 39.15\%, respectively. 
These results demonstrate that the patch pruning component, along with its patch deduplication and regression testing strategies, effectively reduces the ensemble space. 
Combined with the findings in Table~\ref{tab:rq3} (as discussed in RQ3), which show that these components contribute to improved selection effectiveness, we can qualitatively conclude that reducing the ensemble space is beneficial to enhance patch selection performance.

To further quantitatively assess the correlation between the ensemble space and selection effectiveness, we compute the \textit{Pearson's} $r$, \textit{Spearman's} $\rho$, and \textit{Kendall's} $\tau$ between the two metrics (ensemble space and Pass@1).
As shown in Table~\ref{tab:correlation}, the absolute values of \textit{Pearson's} $r$ range from 0.73 to 0.91, \textit{Spearman's} $\rho$ range from 0.80 to 1.00, and \textit{Kendall's} $\tau$ range from 0.67 to 1.00, respectively. 
These results consistently indicate a strong correlation across all three correlation metrics, suggesting that reducing the ensemble space is closely associated with improved selection effectiveness.

%% file: discussion.tex
\section{Discussion}
\label{sec:discussion}

\begin{figure}[t!]
    \centering
    \includegraphics[width=1.0\linewidth]{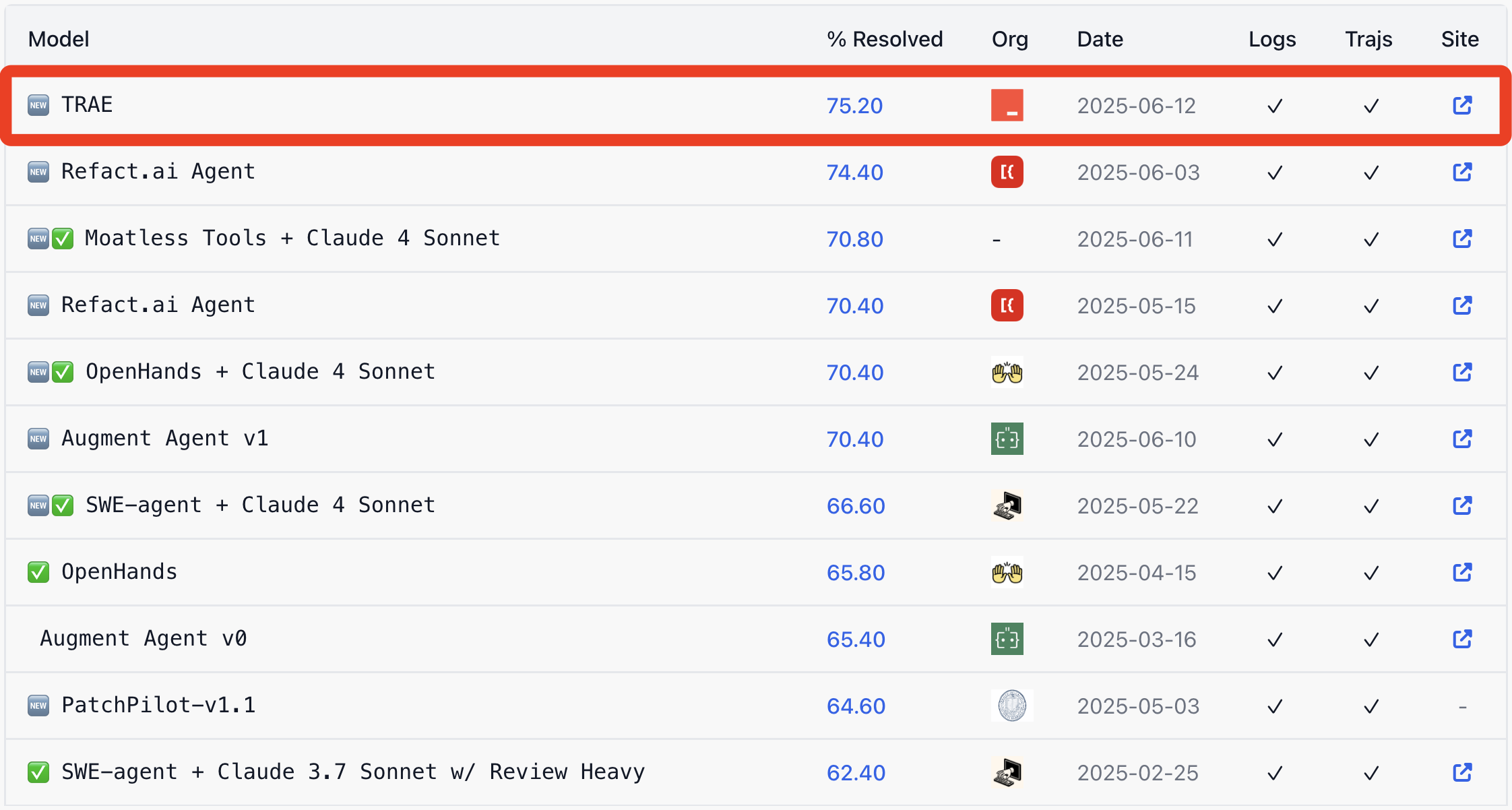}
    \caption{Snapshot of the official SWE-bench Verified leaderboard as of July 2025.}
    \label{fig:leaderboard}
\end{figure}

\subsection{Practical Impact}
Building autonomous agent-based systems for software engineering tasks has emerged as a leading research direction in recent years.
SWE-bench, a widely adopted benchmark, encompasses real-world software engineering tasks such as bug fixing and feature implementation.
As of July 2025, SWE-bench has been downloaded over 500,000 times, with more than 90 agent-based techniques submitted to its official leaderboard.
The SWE-bench leaderboard~\citep{swebleaderboard2025} continuously tracks the state-of-the-art progress in agent-based software engineering systems.
\textbf{As shown in Figure~\ref{fig:leaderboard}, Trae Agent has achieved the first place on the SWE-bench Verified leaderboard, with a notable Pass@1 score of 75.20\%.}
Furthermore, our open-source GitHub repository has attracted over 8,000 stars (as of July 2025), indicating substantial community interest and adoption.
Our findings suggest that integrating ensemble reasoning techniques into existing agent-based systems can significantly improve their ability to resolve complex real-world issues.
We believe that this insight could serve as a guiding principle for the design of future agent-based software engineering frameworks.

\begin{table}[t!]
    \caption{Quality of regression tests in terms of Accuracy ($\uparrow$), Precision ($\uparrow$), Recall ($\uparrow$), F1-Score ($\uparrow$).}
    \label{tab:regression_accuracy}
    \centering
    \tabcolsep=6.8mm
    \normalsize
    \begin{adjustbox}{max width=1.0 \textwidth,center}
        \begin{tabular}{lccccc}
            \toprule
            \midrule
            \textbf{Metric} & \textbf{TP} & \textbf{TN} & \textbf{FP} & \textbf{FN} & \textbf{Total} \\
        	\midrule
            \# Instances & 10,424 & 2,231 & 6,608 & 737 & 20,000 \\
            \midrule
            \multicolumn{5}{l}{Accuracy = (TP+TN)/Total} & 63.28\% \\
            \multicolumn{5}{l}{Precision = TP/(TP+FP)} & 61.20\% \\
            \multicolumn{5}{l}{Recall = TP/(TP+FN)} & 93.40\% \\
            \multicolumn{5}{l}{F1-Score = (2$\times$Precision$\times$Recall)/(Precision+Recall)} & 73.95\% \\
            \midrule
            \bottomrule
            \multicolumn{6}{l}{\footnotesize * TP (true positive): patch passes regression tests and is correct;} \\
            \multicolumn{6}{l}{\footnotesize * TN (true negative): patch does not pass regression tests and incorrect;} \\ 
            \multicolumn{6}{l}{\footnotesize * FP (false positive): patch passes regression tests but is incorrect;} \\
            \multicolumn{6}{l}{\footnotesize * FN (false negative): patch does not pass regression tests but is correct.}
        \end{tabular}
    \end{adjustbox}
\end{table}

\subsection{Quality of Regression Tests}
\label{subsec:influence_of_regression_testing}
In the patch pruning component, the regression tests selected by the tester agent play a critical role in determining the overall effectiveness of \tech{}. 
To evaluate the quality of these regression tests, we assess the consistency between their pruning decisions and the ground-truth results provided by the golden tests in the SWE-bench. 
Specifically, we examine whether the regression tests accurately determine the correctness of the candidate patches.
As shown in Table~\ref{tab:regression_accuracy}, the regression tests achieve an Accuracy of 63.28\%, a Precision of 61.20\%, a Recall of 93.40\%, and an F1-Score of 73.95\%, respectively. 
We further analyze the two types of regression test mispredictions (i.e., FP and FN), which directly affect the effectiveness of patch pruning. 
The overall misprediction rate ($\frac{FP + FN}{Total}$) is 36.73\%. 
Among these, FP cases (i.e., incorrect patches mistakenly passing the regression tests) can be viewed as a conservative patch pruning strategy that preserves erroneous patches for subsequent selection. 
In contrast, FN cases (i.e., correct patches mistakenly discarded) are more detrimental to overall effectiveness, as they reduce the correct candidate patches.
In particular, the FP rate is 33.04\%, while the FN rate accounts for only 3.69\%. 
This indicates that the risk of discarding correct patches is relatively low.

\begin{table}[t!]
    \caption{Influence of the regression testing strategy in terms of the ratios of all-correct ($\uparrow$) and all-incorrect ($\downarrow$) instances.}
    \label{tab:regression_error}
    \centering
    \tabcolsep=1.9mm
    \normalsize
    \begin{adjustbox}{max width=1.0 \textwidth,center}
        \begin{tabular}{cccccc}
            \toprule
            \midrule
            \textbf{Metric} & \textbf{Technique} & \textbf{Gemini 2.5 Pro} & \textbf{Claude 3.7 Sonnet} & \textbf{GPT-4.1} & \textbf{Mixture} \\
        	\midrule
            \multirow{2}{*}{\textbf{All-correct}}  & w/o Regression & 25.80\% & 38.40\% & 26.80\% & 24.20\% \\ 
            & w/ Regression & 40.80\% & 48.00\% & 37.00\% & 37.40\% \\ \midrule

            \multirow{2}{*}{\textbf{All-incorrect}} & w/o Regression & 24.00\% & 21.60\% & 28.20\% & 19.80\% \\ 
            & w/ Regression & 28.60\% & 25.80\% & 32.40\% & 24.60\% \\ 
            \midrule
            \bottomrule
        \end{tabular}
    \end{adjustbox}
\end{table}

In addition, we further investigate the impact of regression tests on instance-level correctness. 
Specifically, each instance is associated with multiple candidate patches. 
If all candidate patches are correct, the instance is guaranteed to be resolved successfully; conversely, if all candidates are incorrect, the instance cannot be resolved. 
Therefore, for ensemble reasoning techniques, it is desirable to maximize the number of all-correct instances and minimize the number of all-incorrect instances.
To this end, we evaluate the ratios of all-correct and all-incorrect instances before and after applying the regression testing strategy. 
As shown in Table~\ref{tab:regression_error}, the regression testing strategy increases the ratio of all-incorrect instances by 4.45\% on average, but more notably increases the ratio of all-correct instances by 12.00\% on average. 
These findings further support the effectiveness of the regression testing strategy and provide an in-depth explanation for its positive impact on the overall performance of \tech{}.

\subsection{Future Work}
\label{subsec:future_work}
\tech{} is the first agent-based ensemble reasoning approach for repository-level issue resolution.
While its effectiveness has been demonstrated through comprehensive empirical studies, several aspects of \tech{} can be further enhanced in future work:
\begin{itemize}[leftmargin=10pt]    
    \item[(1)] \textbf{Enhancing ensemble diversity}: Currently, \tech{} relies on a single coder agent with a high-temperature sampling strategy for diverse patch generation. 
    We plan to integrate multiple issue resolution agents (e.g., Agentless~\citep{xia2024agentless}, Moatless~\citep{moatless2025}, and OpenHands~\citep{wang2025openhands}) to better leverage their complementary strengths. 
    This integration is expected to increase the diversity of candidates, thereby improving the ensemble performance.
    
    \item[(2)] \textbf{Scaling ensemble size}: Our current experiments (as discussed in RQ2) evaluate ensemble sizes ranging from 1 to 10, with results showing a consistent performance improvement of \tech{} as the ensemble size increases. 
    In future work, we plan to investigate the potential performance gains of larger ensemble sizes, while carefully considering cost-efficiency constraints.

    \item[(3)] \textbf{Enhancing patch deduplication}: Recent studies have explored leveraging LLMs for program equivalence detection, such as identifying equivalent mutants~\citep{ma2023lms,tian2024large}. 
    Building on this, we plan to investigate LLM-based patch deduplication techniques to further enhance the overall performance of \tech{}. 
    
\end{itemize}

%% file: threats.tex
\section{Threats and Validity}
\label{sec:threats}
\textbf{Construct Validity.} 
This threat mainly arises from the inherent randomness of LLMs. 
To mitigate this, we conduct a large-scale empirical study and release the replication package for practitioners. 
Moreover, we ensure consistency by repeating all experiments three times in RQ1. 
Notably, the standard deviations of Pass@1 for Augment, Augment w/ Pruning, DeiBase, DeiBase w/ Pruning, and \tech{} are only 0.0052, 0.0039, 0.0038, 0.0044, and 0.0019, respectively, indicating high robustness across runs. 
In addition, a \textit{Wilcoxon Signed-Rank Test}~\citep{wilcoxon1963critical} (at a significance level of 0.05) yields p-values exceeding 0.38 for all comparisons, indicating no statistically significant differences across the three repeated experimental results and further strengthening the reliability of our findings.

\textbf{External Validity.} 
This threat mainly lies in our experimental subjects. 
To address this, we adopt the widely-used SWE-bench benchmark and multiple evaluation metrics commonly used in the issue resolution domain~\citep{zhang2024autocoderover,zhang2024diversity,ruan2025specrover,wang2025openhands}. 
We further evaluate \tech{} against four state-of-the-art ensemble reasoning techniques across three leading LLMs and the widely-used SWE-bench benchmark, ensuring comprehensive comparisons. 
In future work, we plan to extend our evaluation to additional benchmarks and LLMs to further assess the generalizability and robustness of \tech{} across diverse settings.

%% file: related.tex
\section{Related Work}
\label{sec:related}

\subsection{Automatic Software Issue Resolution}
Automatic software issue resolution is a critical task in software engineering and has attracted increasing research interest in recent years. 
Numerous techniques have been proposed, among which agent-based approaches have gained particular prominence.

OpenDevin~\citep{wang2025openhands} (renamed OpenHands) is among the earliest agent-based frameworks, leveraging a planning mechanism based on user requirements and utilizing tools (such as file editors, terminals, and web search engines) to iteratively accomplish complex tasks. 
SWE-agent~\citep{yang2024swe} introduces a custom agent-computer interface, enabling agents to interact with the codebase through operations such as viewing and editing files. 
Moatless~\citep{moatless2025} enhances issue resolution by equipping agents with code search tools and retrieval strategies to identify relevant code locations. 
AutoCodeRover~\citep{zhang2024autocoderover} further refines the code search capability by representing software projects as abstract syntax trees, allowing agents to effectively retrieve contextual information and locate faults. 
Building on AutoCodeRover, SpecRover~\citep{ruan2025specrover} improves the specifications by generating function summaries, and also incorporates the generation of reproduction tests to assist in patch generation. 
Agentless~\citep{xia2024agentless} adopts a standardized operational pipeline comprising fault localization, patch generation, and patch verification, without requiring agents to dynamically decide on future actions or interact with complex external tools.
MarsCode~\citep{liu2024marscode} Agent combines advanced code analysis techniques (i.e., code knowledge graph) with LLM capabilities to provide a systematic process for fault localization, candidate patch generation, and patch validation.

Unlike existing individual patch generation techniques, \tech{} can be seamlessly integrated with them to build more effective ensemble reasoning systems. 
Furthermore, due to its generalizable and modular design, \tech{} potentially provides a promising foundation for advancing ensemble reasoning in broader and more complex software engineering tasks.

\subsection{Ensemble Techniques for LLM}
Ensemble learning~\citep{polikar2012ensemble,sagi2018ensemble} has been widely adopted in machine learning and deep learning models to reduce prediction bias and improve generalization by selecting a consensus solution from multiple models, thereby mitigating the limitations inherent to any single model. 
Traditional ensemble methods (e.g., bagging~\citep{breiman1996bagging}, boosting~\citep{freund1997decision}, and stacking~\citep{dvzeroski2004combining}) have achieved notable success across a variety of domains, including image classification~\citep{chen2019deep}, natural language processing~\citep{zhang2024survey}, and anomaly detection~\citep{han2021gan}.

Recent advances have demonstrated the potential of ensemble techniques to enhance the performance of LLM-based agent systems across a range of tasks. 
In the mathematical reasoning domain, \citet{snell2025scaling} conduct a comprehensive study and introduce the Best-of-N approach, which generates multiple solutions in parallel from a base LLM and employs a reward model to select the highest-scoring solution. 
In the domain of competition-level code generation, ~\citet{li2025s} propose S*, which leverages both the execution feedback from public and LLM-generated test cases along with a clustering-based selection mechanism to guide the final code selection. 
In addition, ~\citet{mahmud2025enhancing} develop EnsLLM, which utilizes CodeBLEU~\citep{ren2020codebleu} for syntactic similarity and CrossHair~\citep{crosshair2025} (a property-based testing tool) for behavior similarity. 
Both similarities are then integrated using a voting mechanism to select the most reliable candidate solution.

In the more complex task of software issue resolution, ~\citet{augmenttop12025} propose Augment, an agent that applies the LLM-as-a-judge paradigm to assess the semantic alignment between the given GitHub issue and a set of candidate patches, ultimately selecting the most relevant patch. 
In addition, ~\citet{zhang2024diversity} introduce DeiBase, which prompts the LLM to generate detailed explanations and confidence scores for each candidate patch, thereby selecting the one with the highest score.
In contrast to existing ensemble techniques, we propose \tech{}, the first agent-based ensemble reasoning approach for repository-level software issue resolution.
\tech{} is designed to enhance LLM-based issue resolution by formulating it as an optimal solution search problem and enhancing LLM performance in issue resolution through a modular agent-based architecture.

%% file: appendix.tex
\newcommand{\cat}[1]{\noindent\textbf{#1}\par}
\newcommand{\catt}[1]{\vspace{0.9\baselineskip}\noindent\textbf{#1}\par}

\section*{Contributors}
\label{sec:contributors}

\begin{minipage}[t]{0.48\textwidth}
\cat{Core Contributors}
Pengfei Gao\textsuperscript{*}\\
Zhao Tian\textsuperscript{*}\\
Xiangxin Meng

\catt{Additional Contributors}
Xinchen Wang\\
Ruida Hu\\
Yuanan Xiao\\
Yizhou Liu\\
Zhao Zhang

\end{minipage}\hfill%
\begin{minipage}[t]{0.48\textwidth}

\cat{Special Thanks}
Junjie Chen\\
Cuiyun Gao\\
Yun Lin\\
Yingfei Xiong

\catt{Project Lead}
Chao Peng

\catt{Team Management}
Xia Liu

\end{minipage}\hfill%

\small \textsuperscript{*} Equal contribution.

\section*{Acknowledgment}

We thank Anthropic for building the anthropic-quickstart project that served as a valuable reference for the tool ecosystem. We are also grateful to all contributors who have supported and improved the \texttt{Trae Agent} open-source project since its release.